\documentclass[lettersize,journal]{IEEEtran}
\usepackage{amsmath,amsfonts}
\usepackage{algorithmic}
\usepackage{algorithm}
\usepackage{array}
\usepackage[caption=false,font=normalsize,labelfont=sf,textfont=sf]{subfig}
\usepackage{textcomp}
\usepackage{stfloats}
\usepackage{url}
\usepackage{verbatim}
\usepackage{graphicx}
\usepackage{cite}
\usepackage{booktabs}
\usepackage{amssymb}
\usepackage{makecell}

\begin{document}

\title{
Generalizable 3D Gaussian Splatting enabled Semantic Coding for Real-Time Immersive Video Communications
}

\author{Dingxi Yang,~\IEEEmembership{Student Member,~IEEE}, Wenqi Guo, Yue Liu,~\IEEEmembership{Member,~IEEE}

Jungong Han,~\IEEEmembership{Senior Member,~IEEE} , Zhijin Qin,~\IEEEmembership{Senior Member,~IEEE}
\thanks{Dingxi Yang and Zhijin Qin are with the Department of
Electronic Engineering, Tsinghua University, Beijing 100084, China, and also with the State Key Laboratory of Space Network and Communications.  (e-mail: ydx23@mails.tsinghua.edu.cn;  qinzhijin@tsinghua.edu.cn)

Wenqi Guo and Jungong Han are with Department of Automation, Tsinghua University, Beijing, China (e-mail: wenqiguo8@gmail.com; jungonghan77@gmail.com)

Yue Liu is with the Faculty of Applied Sciences, Macao Polytechnic University, Macao SAR, China. (e-mail: yue.liu@mpu.edu.mo)}

}



\maketitle

\begin{abstract}
Real-time immersive video communications, particularly high-fidelity 3D telepresence, necessitates a synergistic balance between instantaneous dynamic scene reconstruction and high-efficiency data transmission. While recent advancements in feed-forward 3D Gaussian Splatting (3DGS) have enabled real-time rendering, performing multi-view video coding and 3D reconstruction in a decoupled manner leads to suboptimal compression efficiency and high computational complexity. To address this, we propose GS-SCNet, the first unified end-to-end framework that seamlessly integrates Generalizable 3D Gaussian Splatting reconstruction with a dedicated deep Semantic Coding pipeline. Our architecture is underpinned by two core technical contributions: (i) we introduce a Disparity-Guided Parallel Semantic Codec that exploits epipolar geometric priors to facilitate cross-view contextual interaction via disparity compensation and semantic fusion, thereby enabling real-time parallel processing of stereo streams while significantly enhancing rate-distortion performance, and (ii) we develop a Lightweight Gaussian Parameter Predictor which directly projects decoded semantic latents into 3DGS attributes, obviating the necessity for intermediate pixel-domain reconstruction. By coupling the codec with the task-specific predictor, our framework extracts geometric correlations only once, effectively eliminating the redundant computational bottleneck inherent in conventional decoupled paradigms. Extensive evaluations on both synthetic and real-world human datasets demonstrate that GS-SCNet achieves a superior trade-off across compression efficiency, rendering quality, and real-time performance. Notably,our framework exhibits strong cross-domain generalization and robustness against compression artifacts when applied to out-of-domain real-world data, significantly outperforming conventional decoupled transmission paradigms.
\end{abstract}


\section{Introduction}

IMT-2030 identifies immersive communication as a core scenario for 6G networks. By delivering realistic three-dimensional (3D) visualizations and interactive experiences, it has the potential to fundamentally transform human collaboration and telepresence~\cite{lu2024multi}. Recently, both industry and academia have demonstrated the feasibility of immersive communication through prototype systems. For instance, Google's Project Starline \cite{starline} leverages depth sensing and light-field displays to present remote users as full-scale 3D avatars. Tele-Aloha \cite{tu2024tele} achieves high-authenticity telepresence using sparse RGB cameras. While promising, these systems expose a critical bottleneck for large-scale deployment: satisfying the stringent constraints on bandwidth and end-to-end latency while maintaining high-fidelity, real-time 3D reconstruction.

\begin{figure}[t]
  \centering
  \includegraphics[width=\linewidth]{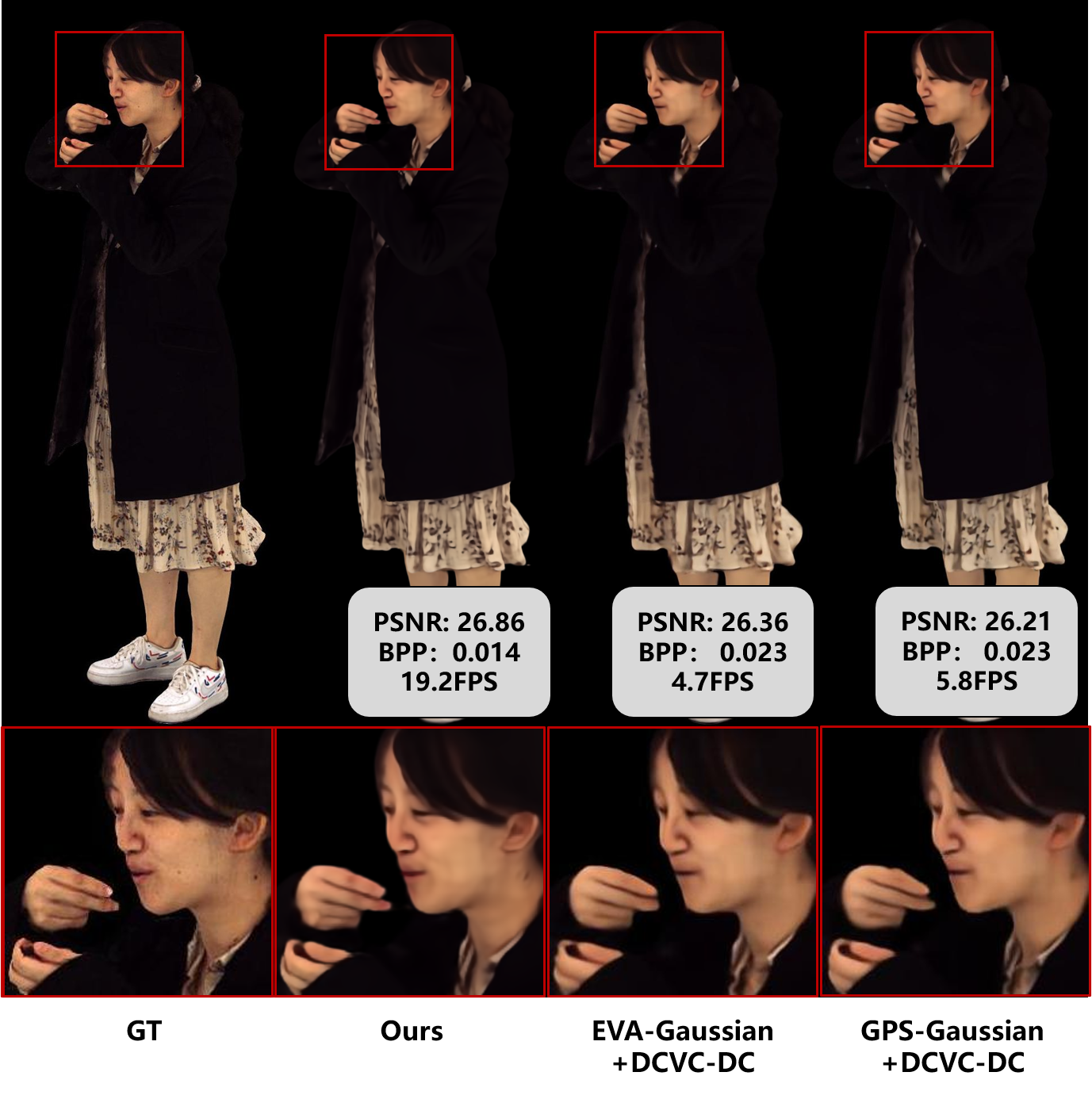}
  \caption{\textbf{Overall performance comparison between our proposed end-to-end GS-SCNet and state-of-the-art decoupled paradigms.} Compared to traditional separated pipelines (e.g., DCVC-DC combined with GPS-Gaussian or EVA-Gaussian), our method significantly outperforms them by achieving higher novel-view rendering quality at lower transmission bitrates, while simultaneously delivering superior real-time inference speed (19.2 FPS).}
  \label{fig:teaser}
\end{figure}

The representation and rendering of dynamic 3D scenes are fundamental to immersive applications. While explicit representations (e.g., meshes and point clouds \cite{huang2022iscom}) require complex acquisition setups, recent advances in neural rendering, such as Neural Radiance Fields (NeRF) \cite{mildenhall2021nerf} and 3D Gaussian Splatting (3DGS) \cite{kerbl20233d}, have enabled high-fidelity novel view synthesis (NVS) directly from sparse RGB images. Specifically, 3DGS discretizes scenes into anisotropic Gaussian ellipsoids, unlocking real-time rendering capabilities. To overcome the time-consuming per-scene optimization required by vanilla 3DGS, recent feed-forward frameworks (e.g., MVSplat \cite{chen2024mvsplat}, GPS-Gaussian \cite{zheng2024gps}, and EVA-Gaussian \cite{hu2024eva}) propose to directly regress Gaussian parameters from multi-view inputs. These methods substantially narrow the gap between 3D reconstruction and real-time inference.

However, as reconstruction fidelity improves, the massive volume of multi-view visual data makes highly efficient compression the primary challenge for practical deployment. Existing telepresence systems typically treat video compression and 3D reconstruction as two isolated tasks. For example, systems like Starline \cite{starline} and Tele-Aloha \cite{tu2024tele} rely on standard High Efficiency Video Coding (HEVC) \cite{H265} or Multi-view HEVC (MV-HEVC) \cite{MVHEVC} to encode each camera stream independently before performing 3D reconstruction at the receiver. Beyond traditional standards, recent learned single-view codecs (e.g., DCVC series \cite{DCVCDC, DCVCRT}) and multiview codecs (e.g., LSVC \cite{chen2022lsvc} and LLSS \cite{hou2024llss}) have been proposed. However, learned multiview codecs often rely on sequential encoding or complex optical flow estimation, introducing prohibitive latency. Alternatively, exploring the direct compression of explicit 3D representations \cite{chen2024hac} struggles with dynamic topologies and bandwidth fluctuations. 

Fundamentally, this decoupled paradigm, where 2D video coding and 3D reconstruction operate in isolation, is highly suboptimal. Since both multi-view video coding and 3DGS reconstruction heavily rely on exploiting inter-view geometric correlations, this artificial separation introduces two critical flaws. 
First, decoupled paradigms inherently suffer from suboptimal coding performance and misaligned optimization objectives. By compressing multi-view streams independently or sequentially, conventional and learned codecs fail to efficiently eliminate cross-view spatial redundancy, causing significant bitrate waste. Furthermore, they strictly optimize for pixel-domain Rate-Distortion (RD), blindly allocating bits to high-frequency noises rather than 3D geometry. Under bandwidth constraints, this misallocation exacerbates compression artifacts and degrades downstream geometry estimation. Ultimately, optimizing purely for RD severely degrades the true objective of immersive communication: the \emph{Rate-Rendering Distortion (RRD)} trade-off, which evaluates the balance between transmission bitrate and synthesized 3D novel-view quality.
Second, this separation causes severe computational redundancy. Because the tasks are isolated, both the multi-view video codec and the subsequent 3D reconstruction network are forced to independently employ high-complexity modules to extract inter-view correlations. This repeated modeling of cross-view geometry incurs unacceptable end-to-end latency for real-time communication.

Consequently, existing paradigms are trapped in a fundamental dilemma: decoupled pixel-domain codecs suffer from 3D-rendering fragility and redundant computational latency, while explicit 3D compression lacks transmission efficiency. To break this impasse, performing semantic-aware compression directly on the multi-view feature streams---guided by shared 3D geometric priors---seamlessly bridges the gap.

\textbf{Motivation and Contributions.} To address the aforementioned bottlenecks, our fundamental perspective is that multi-view video coding and 3D scene reconstruction should not be isolated. Since both processes inherently share the need for inter-view contextual interaction, they must be unified to eliminate the redundant, high-complexity correlation extraction steps and to optimize directly for the RRD objective. Motivated by this, we propose the \textbf{3D Gaussian Splatting-enabled Semantic Codec Network (GS-SCNet)}. By efficiently extracting and utilizing cross-view correlations through a shared disparity-guided pipeline, GS-SCNet realizes a low-latency, highly efficient joint coding and feed-forward reconstruction paradigm.

Unlike existing pipelines that sequentially compress pixels and then estimate 3D parameters, GS-SCNet directly extracts and compresses semantic features from stereo RGB inputs. We design a Disparity-Guided Parallel Semantic Codec that leverages epipolar geometry to enable parallel encoding of left and right views. By introducing a novel disparity compensation and semantic fusion mechanism, our codec effectively eliminates inter-view redundancy at the feature level without relying on latency-heavy motion estimation. Subsequently, a lightweight predictor regresses all 3DGS parameters directly from the decoded semantic features in a single feed-forward pass, enabling high-fidelity rendering. 

In summary, the main contributions of this work are as follows:
\begin{enumerate}
   
    \item To resolve the suboptimality of isolated compression and rendering tasks, we propose GS-SCNet, a novel unified framework that deeply couples semantic multi-view video coding with generalizable 3D Gaussian Splatting. This shifts the optimization paradigm from traditional pixel-domain RD to RRD.

    \item To exploit cross-view geometric correlations without introducing the prohibitive latency of sequential coding, we design a disparity-guided cross-view semantic autoencoder. By explicitly utilizing disparity for semantic fusion, the network achieves parallel stereo encoding, simultaneously suppressing redundancy and ensuring real-time performance.

    \item To address the severe computational bottleneck at the receiver, we develop a lightweight Gaussian-parameter prediction module. By directly deriving 3D splatting attributes from the decoder's semantic feature maps and decoded disparity, it inherently prevents the repeated, high-complexity extraction of cross-view geometry that plagues decoupled pipelines, achieving ultra-low inference latency while explicitly preserving geometric priors.

    \item Extensive evaluations on real-world datasets demonstrate that GS-SCNet effectively overcomes the vulnerability of decoupled methods to compression noise. Our system exhibits exceptional cross-domain generalization and robustness, maintaining consistent geometric alignment and rendering quality where decoupled paradigms struggle with scale ambiguity and cross-view fusion artifacts.
\end{enumerate}


A preliminary version of this work appeared in our conference paper on immersive video communications based on semantic coding and generalizable 3D Gaussian Splatting\cite{DX}. In this journal extension, we substantially enhance both the system design and the empirical evaluations. Specifically, we upgrade the semantic compression pipeline by integrating an advanced context-adaptive entropy model to further boost coding efficiency. Moreover, we significantly expand the experimental scope by introducing more representative decoupled baselines (e.g. EVA-Gaussian\cite{hu2024eva}) and conducting rigorous cross-domain evaluations on an out-of-domain real-world dataset (DNA-Rendering\cite{cheng2023dna}) to validate our framework's robustness and generalization capabilities. 

The remainder of this paper is organized as follows. Section II reviews related work on neural video compression and feed-forward 3D Gaussian Splatting. Section III introduces the overall framework and details the architecture of the proposed GS-SCNet. Section IV presents the experimental setup, extensive comparisons with state-of-the-art methods, and cross-domain generalization analysis. Section V concludes the paper.

\section{Related Work}

\textbf{Neural Video Codecs.} Since the introduction of the end-to-end optimized image compression model presented in \cite{balle2016end} and its extension with a hyperprior\cite{balle2018variational}, neural image codecs (NIC) have witnessed rapid development \cite{minnen2018joint} \cite{hu2020coarse}\cite{he2022elic}. The standardization of NIC is actively underway\cite{ascenso2023jpeg}.
Building on the architectural principles and successes of NICs, researchers have begun extending similar deep learning paradigms to the neural video compression (NVC) domain.
Deep learning-based video coding typically integrates techniques such as motion estimation, motion compensation, nonlinear transformation, and entropy coding. For example, DVC\cite{DVC} employs these components to construct the first end-to-end neural network-based video residual coding framework. It predicts the current frame using optical flow estimation and motion compensation, and subsequently encodes both the optical flow and the residual. Considering the similarity of motion vectors between adjacent frames, several works \cite{rippel2021elf}\cite{pourreza2023boosting} further reduce redundancy through optical flow-based motion prediction. FVC \cite{hu2021fvc} shifts encoding operations such as motion compensation and residual computation from the pixel domain to the feature domain, and leverages deformable convolution to perform efficient motion compensation, thereby enhancing coding performance. However, these residual coding methods always rely on simple subtraction to compress temporal redundancy, which can lead to suboptimal performance. To address this, conditional coding methods have been proposed, which reduce redundancy by learning inter-frame contextual information, thereby enhancing the flexibility of the coding framework \cite{li2021deep}. Building on this idea, DCVC-TCM \cite{sheng2022temporal} improves compression performance by introducing feature propagation, while DCVC-HEM\cite{li2022hybrid} proposes a hybrid spatial-temporal entropy model that incorporates latent priors and dual spatial priors to effectively capture spatial and temporal dependencies in video. Furthermore, DCVC-DC \cite{DCVCDC} extracts diverse contextual features and introduces an entropy model based on quadtree-based feature partitioning, outperforming state-of-the-art traditional codecs in both RGB and YUV420 color spaces.

After achieving notable improvements in compression efficiency, research efforts have shifted towards enhancing the real-time performance and practical deployability of NVC. However, most existing approaches still struggle to achieve real-time compression for high-resolution videos\cite{rippel2021elf}\cite{tian2023towards}. The recent work DCVC-RT\cite{DCVCRT} enables real-time encoding and decoding of 1080p videos by removing motion-related modules and employing latent representations at a single-resolution.

Disparity has long been used to exploit inter-view correlation in multiview coding, from early disparity-compensated schemes~\cite{perkins2002data} to standards such as multiview video coding (MVC)~\cite{MVC} and MV-HEVC~\cite{MVHEVC}. Building on this idea, LSVC~\cite{chen2022lsvc} presents the first end-to-end optimized stereoscopic codec that sequentially encodes views using motion- and disparity-compensated references, improving rate–distortion over MV-HEVC but incurring high latency. LLSS~\cite{hou2024llss} instead processes the two views in parallel with implicit disparity estimation, reducing inference latency via a parallel autoencoder, yet still falling short of real-time at high resolutions. Moreover, these learning-based methods are mostly developed on datasets with relatively small disparities, and their performance degrades for the large-disparity stereo typical of our setting. Crucially, these existing stereo/multiview codecs optimize purely for pixel-domain fidelity (i.e., Rate-Distortion). In contrast, our cross-view modeling explicitly fuses semantic features guided by epipolar geometry, optimizing specifically for the downstream 3D rendering task (Rate-Rendering Distortion). 

Furthermore, recent works have begun to explore the direct compression of 3D representations for telepresence, such as the hierarchical compression of dynamic 3D avatars (e.g., HGC-Avatar \cite{tang2025hgc}). However, directly transmitting explicit 3D parameters is often fragile to dynamic topology changes and strict bandwidth fluctuations. Our approach circumvents this by compressing robust 2D semantic features and lifting them to 3D representations at the receiver, bridging the gap between efficient 2D coding and 3D rendering.

\textbf{Feed-Forward 3DGS.} 3DGS employs a tile-based differentiable rasterizer to enable real-time rendering\cite{kerbl20233d}. Recent advances in feedforward 3DGS investigate generalizable 3DGS representation reconstruction from sparse views, aiming to address the limitation of the vanilla 3DGS framework, which requires per-scene optimization. Splatter Image\cite{szymanowicz2024splatter} enables rapid reconstruction of 3DGS representations from a single image. However, the rendering quality is constrained by the ill-posed nature of monocular 3D reconstruction. PixelSplat\cite{charatan2024pixelsplat} and MVSplat\cite{chen2024mvsplat} perform 3D reconstruction for general scenes by regressing Gaussian parameters from multi-view inputs. GPS-Gaussian \cite{zheng2024gps} integrates stereo matching techniques to reconstruct 3D human representations in real time from a pair of rectified stereo images. Its extended version, GPS-Gaussian++\cite{zhou2025gps}, further generalizes this approach to handle more complex human-centered scene data. Our model also employs stereo matching to estimate disparity, which is subsequently used to reduce inter-view redundancy and to compute the Gaussian centers. 

Furthermore, recent pioneering works like GHG \cite{kwon2024generalizable} and EVA-Gaussian \cite{hu2024eva} have achieved impressive sparse-view synthesis for dynamic humans by incorporating human template priors (e.g., SMPL) or efficient global attention mechanisms. However, regarding their suitability for real-time telepresence, these pipelines typically assume uncompressed, pristine RGB inputs, completely ignoring transmission bandwidth bottlenecks. Additionally, their reliance on explicit human mesh fitting or dense cross-view attention operations significantly increases end-to-end latency. Distinct from these purely reconstruction-focused pipelines, our GS-SCNet uniquely couples a lightweight Gaussian predictor with a semantic codec, tailored specifically to meet the strict Rate-Rendering Distortion and latency constraints of real-time communication.

\begin{figure*}[t!]
\centerline{\includegraphics[width=1.0\linewidth]{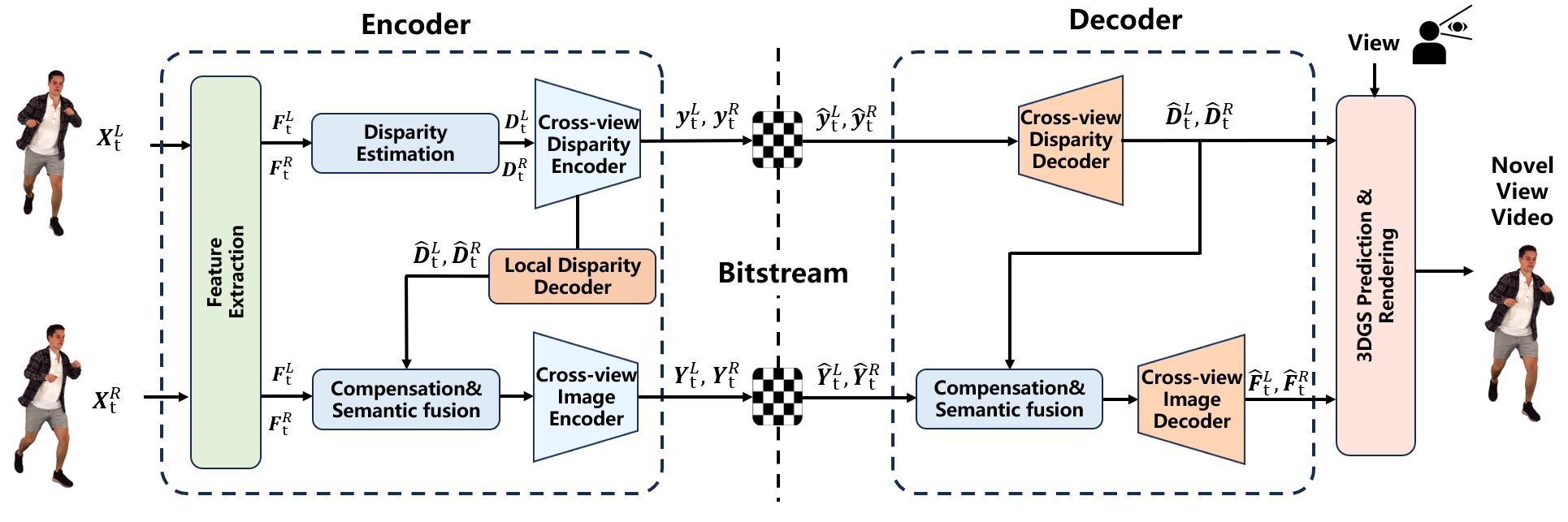}}
\caption{\textbf{Framework of the proposed GS-SCNet.} Taking cross-view (left and right) video streams as input, our model employs two fully parallel, weight-shared branches. At the Encoder, each frame undergoes feature extraction and disparity estimation. The estimated disparities are then compressed by the Cross-view Disparity Encoder. Guided by the locally decoded disparities, the semantic features of the two views undergo contextual interaction via the \textit{Compensation \& Semantic Fusion} module to effectively suppress inter-view redundancy. Subsequently, the fused representations are compressed by the Cross-view Image Encoder into a bitstream. At the Decoder, the cross-view disparities are first reconstructed. Guided by these disparities, the image semantic features ($\hat{F}_t^L, \hat{F}_t^R$) are fully restored through an identical semantic fusion and decoding process. Finally, lightweight predictors directly infer 3D Gaussian parameters from these decoded semantic features. Novel-view human videos are rendered using the predicted 3D Gaussians.}
\label{fig1}
\end{figure*}

\section{Method}

\subsection{Overview}


Figure \ref{fig1} illustrates the overall architecture of the proposed GS-SCNet for real-time immersive communication. Given synchronized left and right video frames captured by calibrated sparse cameras, the transmitter encodes the stereo inputs into compact disparity and semantic bitstreams. At the receiver, these bitstreams are decoded into geometry-aware semantic representations, which are directly converted into 3D Gaussian parameters for novel-view rendering. Therefore, the goal is not merely to reconstruct the transmitted stereo frames in the pixel domain, but to preserve the geometry and appearance information required for high-quality target-view rendering at the receiver. This formulation contrasts with conventional decoupled pipelines, where multi-view RGB streams are first compressed and decoded by a video codec, and the decoded frames are then fed into a separate 3DGS reconstruction network. Since the reconstruction stage operates only on pixel-domain frames, it has to infer cross-view geometric correlations again from the compressed images, leading to additional computation and increased sensitivity to compression artifacts. GS-SCNet instead decodes disparity-guided semantic features and directly uses them for lightweight Gaussian parameter prediction. The geometric correspondences established during semantic coding are therefore reused for 3DGS parameter prediction, avoiding redundant cross-view correlation modeling. To implement this communication-to-rendering pipeline, GS-SCNet employs a dual-branch, weight-shared architecture to process the left and right views in parallel. The entire pipeline proceeds through three core stages: 1) cross-view disparity estimation and compression to capture geometric priors, 2) disparity-guided semantic feature coding to eliminate inter-view spatial redundancy, and 3) lightweight 3DGS parameter prediction directly from decoded semantic features.

At each time step $t$, the left and right views are processed concurrently. Given the input frames $\mathbf{X}_t^L$ and $\mathbf{X}_t^R$, the network first estimates the corresponding disparities $\mathbf{D}_t^L$ and $\mathbf{D}_t^R$. To facilitate geometry-aware decoding at the receiver, these disparities are compressed by a dedicated cross-view disparity autoencoder:
\begin{equation}
\mathbf{y}_t^V = \mathcal{E}_{\text{disp}}\big(\mathbf{D}_t^V, \hat{\mathbf{f}}_{t-1}^V \big), 
\quad V \in \{L,R\},
\end{equation}
where $\mathcal{E}_{\text{disp}}(\cdot)$ denotes the disparity encoder conditioned on the previously decoded disparity features $\hat{\mathbf{f}}_{t-1}^V$. The latent codes $\mathbf{y}_t^L, \mathbf{y}_t^R$ are quantized into $\hat{\mathbf{y}}_t^L, \hat{\mathbf{y}}_t^R$ and entropy-coded for transmission. Subsequently, the disparity decoder $\mathcal{D}_{\text{disp}}(\cdot)$ reconstructs the disparities as $\hat{\mathbf{D}}_t^L, \hat{\mathbf{D}}_t^R$:
\begin{equation}
\hat{\mathbf{D}}_t^V = \mathcal{D}_{\text{disp}}(\hat{\mathbf{y}}_t^V), 
\quad V \in \{L,R\}.
\end{equation}

To effectively suppress inter-view redundancy, the intermediate image semantic features of the two views ($\mathbf{F}_t^L, \mathbf{F}_t^R$) undergo a cross-view contextual interaction. Guided by the reconstructed disparities, the features are warped and fused to generate disparity-aware representations:
\begin{align}
\overline{\mathbf{F}}_t^L &= \Phi\!\left(\mathbf{F}_t^L, \text{warp}(\mathbf{F}_t^R, \hat{\mathbf{D}}_t^L)\right), \nonumber \\
\overline{\mathbf{F}}_t^R &= \Phi\!\left(\mathbf{F}_t^R, \text{warp}(\mathbf{F}_t^L, \hat{\mathbf{D}}_t^R)\right),
\end{align}
where $\text{warp}(\cdot)$ denotes the disparity-guided spatial warping operation that aligns the opposite view's features to the current view, and $\Phi(\cdot)$ represents the semantic fusion function.

Subsequently, based on these interacting features and the temporal context from previous frames ($\hat{\mathbf{F}}_{t-1}^V$), the cross-view image encoder produces the final latent representations:
\begin{equation}
\mathbf{Y}_t^V = \mathcal{E}_{\text{img}}\big(\mathbf{X}_t^V, \hat{\mathbf{F}}_{t-1}^V, \hat{\mathbf{D}}_t^V \big), 
\quad V \in \{L,R\}.
\end{equation}
These representations are quantized, entropy-coded, and transmitted. At the decoder, an identical semantic fusion process is mirrored to fully restore the image semantic features $\hat{\mathbf{F}}_t^L, \hat{\mathbf{F}}_t^R$.


Finally, utilizing the decoded disparity features $\hat{\mathbf{f}}_t^V$ and the reconstructed image semantic features $\hat{\mathbf{F}}_t^V$, the lightweight predictor $\Psi(\cdot)$ directly infers the complete set of 3DGS parameters (including scale $\mathbf{S}_t^V$, rotation $\mathbf{R}_t^V$, opacity $\mathbf{O}_t^V$, color residual $\mathbf{C}_{r,t}^V$, and depth residual $\mathbf{Z}_{r,t}^V$) in a single feed-forward pass:
\begin{equation}
\{\mathbf{S}_t^V, \mathbf{R}_t^V, \mathbf{O}_t^V, \mathbf{C}_{r,t}^V, \mathbf{Z}_{r,t}^V\} = \Psi\!\left(\hat{\mathbf{f}}_t^V, \hat{\mathbf{F}}_t^V, \mathcal{T}(\hat{\mathbf{D}}_t^V)\right), 
\end{equation}
where $V \in \{L,R\}$ indexes the left and right views, and $\mathcal{T}(\cdot)$ transforms disparity to depth. These parameters are then unprojected into 3D space to render novel-view human videos via differentiable splatting. By leveraging the shared geometric priors encoded in the semantic features, this single feed-forward pass effectively avoids the redundant cross-view correlation modeling typical of decoupled methods.

\subsection{Disparity Estimation}

It is important to note that for rectified stereo image pairs aligned horizontally, the corresponding pixels in the left and right views lie on the same horizontal line. Therefore, the disparity can be predicted based on pixel-wise displacement. The disparity, representing the horizontal shift between matched keypoints, can be converted into depth information using known camera parameters such as focal length and baseline distance.

Traditional stereo matching methods estimate disparity by transforming image coordinates into world coordinates and performing cost volume construction and disparity regression. For instance, 3DGS requires accurate camera alignment to ensure the precision of central point estimation; otherwise, image degradation during rendering may significantly reduce quality. Furthermore, the disparity encodes relative spatial information between views, which is crucial for maintaining consistency and compression efficiency during view synthesis.

The disparity estimation module adopts the implementation method proposed in GPS-Gaussian\cite{zheng2024gps}, which constructs a cost volume and updates it iteratively to avoid the computational overhead associated with slow 3D convolutions. To further improve the model's real-time performance, the feature extraction from frames does not employ the conventional approach of progressive downsampling via standard convolutional layers. Instead, the frames are directly downsampled by a factor of 8 using a Patchify module, followed by feature extraction using depthwise separable convolution modules. The disparity estimation process can be expressed as
\begin{equation}
\left\langle \mathbf{D}_t^L, \mathbf{D}_t^R \right\rangle = \phi_{\text{disparity}} \left( f_d \left( \mathbf{X}_t^L, \mathbf{X}_t^R \right), \mathbf{K}_t^L, \mathbf{K}_t^R \right),
\end{equation}
where $f_d$ denotes the feature extractor for the disparity estimation module, $\phi_{\text{disparity}}$ represents the disparity estimation network, and $\mathbf{K}_t^L$, $\mathbf{K}_t^R$ denote the camera intrinsics for the left and right views, respectively.

\subsection{Cross-View Parallel Semantic Feature Autoencoders}

\begin{figure*}[tb]
\centerline{\includegraphics[width=1.0\linewidth]{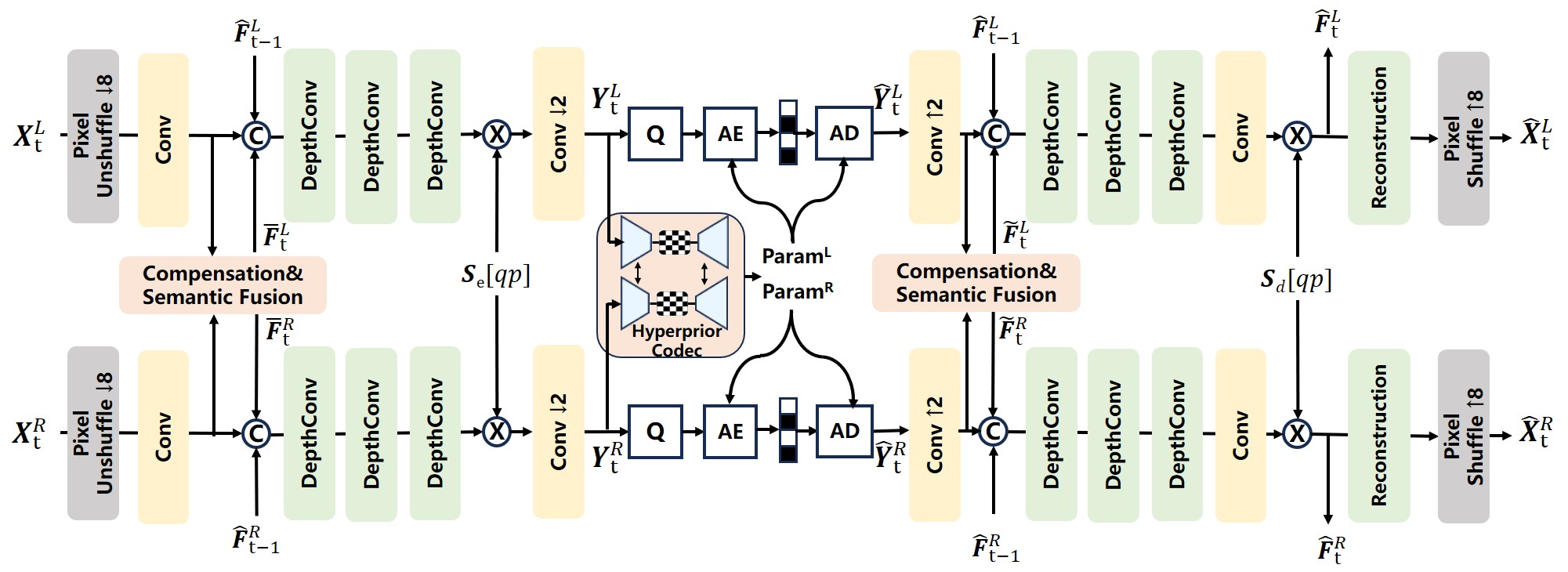}}
\caption{\textbf{Framework of the proposed Cross-View Parallel Image Semantic Feature Autoencoder.} DepthConv, Q, AE and AD represent depthwise convolution block, quantization, arithmetic encoder and decoder, respectively. 
}
\label{fig2}
\end{figure*}

As shown in Fig.~\ref{fig1}, our method employs two cross-view parallel semantic feature autoencoders to successively encode disparity and image features. Each view (left and right) employs a separate encoder–decoder branch, but all parameters are fully shared to maintain consistency and reduce the model size. The structure of the cross-view parallel semantic feature autoencoder for images is depicted in Fig.~\ref{fig2}, while the disparity autoencoder follows a similar design. The only difference lies in the decoder stage: since opposite-view disparity features are unavailable at the receiver, the disparity decoder must reconstruct $\hat{\mathbf{D}}_t^V$ independently without cross-view context.  

\textbf{Semantic Encoder.}  
Each autoencoder begins with a \textit{Patchify} module that downsamples the input disparity or image by a factor of 8. This provides a compact representation and reduces the computational load. The downsampled inputs are then passed through a series of cascaded depthwise separable convolution blocks, denoted as DepthConv, which replace standard convolutions to achieve lower complexity while preserving representation capacity~\cite{chollet2017xception}. The output features are subsequently modulated by a quantization step size, generating highly compressible latent representations. 
The encoding process can be written as
\begin{align}
\mathbf{y}_t^V &= \mathcal{E}_{\text{disp}}\!\left(\mathbf{D}_t^V, \hat{\mathbf{f}}_{t-1}^V\right), \nonumber \\
\mathbf{Y}_t^V &= \mathcal{E}_{\text{img}}\!\left(\mathbf{X}_t^V, \hat{\mathbf{F}}_{t-1}^V, \hat{\mathbf{D}}_t^V\right),
\end{align}
where $\mathbf{Y}_t^V$ and $\mathbf{y}_t^V$ are the latent features for images and disparity, respectively, and $V \in \{L,R\}$. 
It is worth noting that the image and disparity encoders cannot be simply merged. Disparity must be explicitly estimated and decoded first, as the reconstructed disparity $\hat{\mathbf{D}}_t^V$ is indispensable for conditioning the image encoder to perform accurate cross-view geometric warping and feature alignment.

\textbf{Quantization and Entropy Model.} 
To prepare the latent features for transmission, they are quantized and entropy coded. In practice, the quantizer $Q(\cdot)$ is non-differentiable. Therefore, during training, we approximate quantization by adding uniform noise:
\begin{equation}
\tilde{\mathbf{y}}_t^V = \mathbf{y}_t^V + \mathcal{U}\!\left(-\tfrac{1}{2}, \tfrac{1}{2}\right),
\end{equation}
while during inference we apply rounding:
\begin{equation}
\hat{\mathbf{y}}_t^V = \text{round}(\mathbf{y}_t^V).
\end{equation}

To accurately capture the probability distribution of the latent representations, we employ an advanced context-adaptive entropy model inspired by DCVC-DC~\cite{DCVCDC}. Specifically, we adopt its quadtree-based partition scheme, which efficiently increases spatial context diversity while supporting parallel encoding.Let $\mathbf{z}_t^V$ denote the hyperprior representation, and $\hat{\mathbf{z}}_t^V$ be its quantized version. The conditional distribution of the quantized latent features is modeled as a Gaussian distribution:

\begin{equation}
p(\hat{\mathbf{y}}_t^V|\hat{\mathbf{z}}_t^V, \mathbf{c}_t^V) = \prod_{i} \mathcal{N}\!\big(\hat{y}_{t,i}^V \,;\, \mu_{t,i}^V(\hat{\mathbf{z}}_t^V, \mathbf{c}_{t,i}^V), {\sigma_{t,i}^V}^2(\hat{\mathbf{z}}_t^V, \mathbf{c}_{t,i}^V)\big),
\end{equation}
where $\mathbf{c}_{t,i}^V$ represents the local context for the $i$-th element, derived from the quadtree-based spatial neighbors and the previously decoded temporal features. The parameters $\mu_{t,i}^V$ and ${\sigma_{t,i}^V}^2$ are the mean and variance predicted by the entropy parameter network.

The entropy model provides an estimate of the bit rate. The expected number of bits per pixel (bpp) is given by integrating over both stereo views:
\begin{equation}
R_t = \frac{1}{2N} \sum_{V \in \{L,R\}} \sum_{i=1}^{N} -\log_2 p(\hat{y}_{t,i}^V|\hat{\mathbf{z}}_t^V, \mathbf{c}_{t,i}^V),
\end{equation}
where $N$ denotes the total number of pixels in a single view, and $2N$ represents the combined pixels of the stereo pair. Furthermore, in our framework, the hyperprior generation is explicitly extended with a parallel cross-view mode. Instead of relying solely on single-view information, disparity-compensated features from the opposite branch are fused to provide additional cross-view context for the entropy modeling. This disparity-guided semantic fusion significantly enhances the probability estimation accuracy, effectively reducing the final transmission bitrate.

\textbf{Semantic Decoder.}  
At the receiver side, the latent representations $\hat{\mathbf{Y}}_t^V$ and $\hat{\mathbf{Z}}_t^V$ are entropy decoded and then passed through decoders.
An image decoder, denoted as $\mathcal{D}_{\text{img}}$, is employed to decode the image semantic features as follows:
\begin{equation}
\hat{\mathbf{F}}_t^V = \mathcal{D}_{\text{img}}\!\left(\hat{\mathbf{Y}}_t^V, \hat{\mathbf{F}}_{t-1}^V, \hat{\mathbf{D}}_t^V\right),
\end{equation}
and for the disparity maps, the decoder is applied as
\begin{equation}
\hat{\mathbf{D}}_t^V = \mathcal{D}_{\text{disp}}\!\left(\hat{\mathbf{y}}_t^V\right).
\end{equation}
Each decoder consists of multiple depthwise separable convolution layers, followed by a reconstruction module that restores the disparity or image to the original resolution using pixel shuffle layers.  

\textbf{Contextual Conditioning.}  
Both encoders and decoders are conditioned on semantic features from the previous frame, inspired by~\cite{DCVCRT}, to capture temporal dependencies. Unlike traditional motion-compensation-based codecs, we do not perform explicit motion estimation, which would significantly slow down inference. By eliminating motion vector modules, more parameters can be allocated to semantic feature extraction and entropy modeling, resulting in a stronger compression model without sacrificing real-time performance.  

To enable semantic interaction across views, disparity compensation and semantic fusion are performed to generate cross-view contextual representations. These mechanisms are integrated both in the encoder and in the hyperprior network, so that inter-view redundancy is effectively reduced. However, to avoid duplication, the detailed formulations of cross-view compensation and fusion are deferred to the subsequent section.

\subsection{Disparity Compensation and Semantic Fusion}

In rectified stereo video, disparity plays a role analogous to motion vectors in temporal prediction, but along the spatial dimension: it specifies how pixels in one view correspond to pixels in the other. Leveraging these correspondences, large portions of one view can be inferred from its counterpart, which substantially reduces inter-view redundancy. At the same time, disparity-guided warping offers a principled way to handle occlusions: pixels that are visible in only one view can be projected to the other view to reconstruct missing content. This disparity-compensated cross-view prediction serves as the backbone of our semantic interaction mechanism between stereo branches.

Concretely, given the left-view features $\mathbf{F}_t^L$ and the reconstructed right-view disparity $\hat{\mathbf{D}}_t^R$, the predicted right-view representation $\widetilde{\mathbf{F}}_t^R$ can be obtained through disparity-based warping as
\begin{equation}
\widetilde{\mathbf{F}}_t^R = \text{warp}(\mathbf{F}_t^L, \hat{\mathbf{D}}_t^R),
\end{equation}
where $\widetilde{\mathbf{F}}_t^R$ denotes the synthesized right-view feature from the left view. Similarly, the left-view disparity prediction for consistency checking can be written as
\begin{equation}
\widetilde{\mathbf{D}}_t^L = \text{warp}(-\hat{\mathbf{D}}_t^R, \hat{\mathbf{D}}_t^L).
\end{equation}
This formulation establishes bidirectional pixel correspondences between the two views and provides an explicit model for cross-view correlations. To strictly avoid encoder-decoder drift, all warping operations are guided by the reconstructed disparities $\hat{\mathbf{D}}_t$.

However, direct disparity compensation inevitably suffers from occlusion artifacts and local geometric mismatches, especially under large baseline angles. To alleviate these issues, we introduce a disparity-aware confidence weighting mechanism for semantic feature fusion. Intuitively, pixels that exhibit consistent disparities across the two views should receive higher confidence, while inconsistent pixels (likely occluded or misaligned) should be down-weighted. The confidence maps for the left and right views are defined as
\begin{align}
\mathbf{W}_t^L &= \exp\!\left(-\frac{(\hat{\mathbf{D}}_t^L - \text{warp}(-\hat{\mathbf{D}}_t^R, \hat{\mathbf{D}}_t^L))^2}{S^2}\right), \nonumber\\
\mathbf{W}_t^R &= \exp\!\left(-\frac{(\hat{\mathbf{D}}_t^R - \text{warp}(-\hat{\mathbf{D}}_t^L, \hat{\mathbf{D}}_t^R))^2}{S^2}\right),
\end{align}
where $\mathbf{W}_t^L$ and $\mathbf{W}_t^R$ measure the semantic correspondence confidence. Here, $S$ is a learnable scale parameter that controls the width of the Gaussian kernel. A larger $S$ allows smoother weighting with tolerance to disparity mismatches, while a smaller $S$ enforces stricter consistency, making the weighting more selective.  

Finally, the fused image semantic features $\widetilde{\mathbf{F}}_t^L$ and $\widetilde{\mathbf{F}}_t^R$ are obtained by combining original and warped features using these confidence maps:
\begin{align}
\widetilde{\mathbf{F}}_t^L &= \mathbf{W}_t^L \odot \text{warp}(\mathbf{F}_t^R, \hat{\mathbf{D}}_t^L) + (1 - \mathbf{W}_t^L) \odot \mathbf{F}_t^L, \nonumber \\
\widetilde{\mathbf{F}}_t^R &= \mathbf{W}_t^R \odot \text{warp}(\mathbf{F}_t^L, \hat{\mathbf{D}}_t^R) + (1 - \mathbf{W}_t^R) \odot \mathbf{F}_t^R,
\end{align}
where $\odot$ represents element-wise multiplication. Through this weighted fusion, reliable correspondences are reinforced, while noisy or occluded regions are suppressed.  

This disparity compensation and semantic fusion module thus ensures that the fused features $\widetilde{\mathbf{F}}_t^L$ and $\widetilde{\mathbf{F}}_t^R$ carry complementary contextual cues from both views, enhancing semantic consistency and robustness for downstream Gaussian parameter prediction and rendering.

\subsection{Gaussian Parameter Prediction}

\begin{figure}[tb]
\centerline{\includegraphics[width=0.9\linewidth]{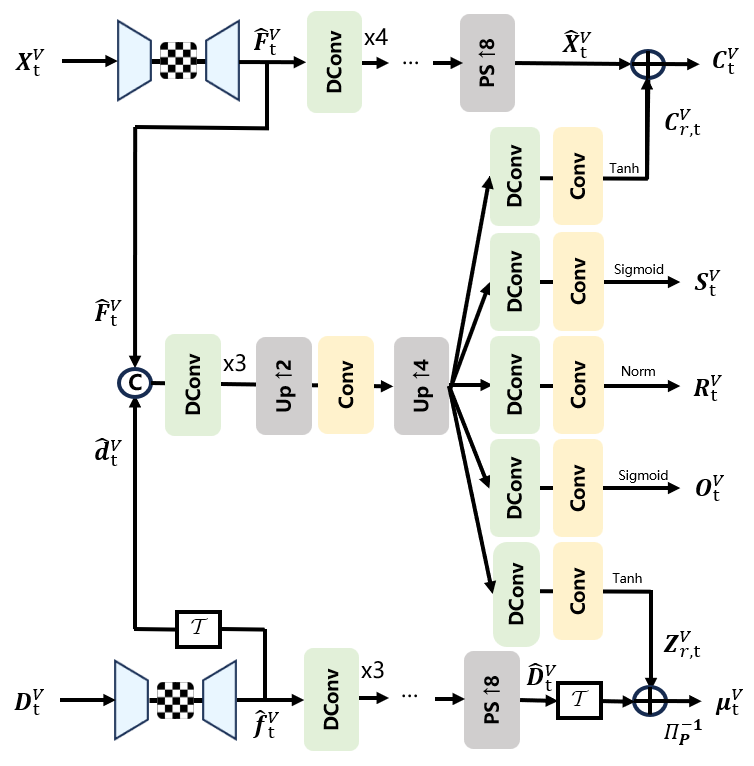}}
\caption{\textbf{The detailed structure of the Gaussian parameter predictor.} DConv, PS and Up represent depth-wise convolution block, pixel shuffle and bilinear upsampling, respectively. A linear transformation $\mathcal{T}$ from disparity to depth is applied to the decoded disparity features $\hat{\mathbf{f}}_t^V$ to obtain the depth semantic features $\hat{\mathbf{d}}_t^V$. The features $\hat{\mathbf{d}}_t^V$ are concatenated with the decoded image semantic features $\hat{\mathbf{F}}_t^V$ to predict Gaussian parameters, including scale $\mathbf{S}_t^V$, rotation $\mathbf{R}_t^V$, opacity $\mathbf{O}_t^V$, color residual $\mathbf{C}_{r,t}^V$, and depth residual $\mathbf{Z}_{r,t}^V$.}
\label{fig3}
\end{figure}

The architecture of the Gaussian-parameter prediction module is illustrated in Fig.~\ref{fig3}.

\textbf{Covariance Matrix and Opacity.}
The decoded semantic features from the left and right views, denoted as $\hat{\mathbf{f}}^L_t$, $\hat{\mathbf{f}}^R_t$, $\hat{\mathbf{F}}^L_t$, and $\hat{\mathbf{F}}^R_t$, are first passed through a shared feature-processing module and then into three lightweight prediction heads to estimate the 3D Gaussian covariance (parameterized by scale and rotation) and opacity:
\begin{align}
\mathbf{S}_t^V &= S_{\max} \cdot \mathrm{Sigmoid}\left( h_s\big(f_p\big(\mathcal{T}(\hat{\mathbf{f}}_t^V), \hat{\mathbf{F}}_t^V\big)\big)\right), \nonumber \\
\mathbf{R}_t^V &= \mathrm{Norm}\left( h_r\big(f_p\big(\mathcal{T}(\hat{\mathbf{f}}_t^V), \hat{\mathbf{F}}_t^V\big)\big)\right), \nonumber \\
\mathbf{O}_t^V &= \mathrm{Sigmoid}\left( h_o\big(f_p\big(\mathcal{T}(\hat{\mathbf{f}}_t^V), \hat{\mathbf{F}}_t^V\big)\big)\right),
\end{align}
\noindent where $V \in \{L, R\}$ indexes the left and right views. Because disparity values and image intensities lie on very different numeric scales, we first map the disparity-domain semantic features to the depth domain via the linear transformation $\mathcal{T}$. Empirically, performing feature aggregation after this disparity–depth conversion leads to faster and more stable convergence. The function $f_p(\cdot)$ denotes the shared module that fuses image and semantic information, while $h_s$, $h_r$, and $h_o$ are lightweight heads that output the scale, rotation, and opacity, respectively. Each head consists of a depthwise separable convolution followed by a standard convolution layer, with activation and normalization schemes chosen according to the valid range of each Gaussian parameter.

\textbf{Coarse Depth and Color.}
After the disparity map is decoded, depth values are obtained through a linear projection based on the known camera intrinsic parameters. Following prior work which observes that human surfaces predominantly exhibit diffuse reflectance~\cite{zheng2024gps}, we adopt a zeroth-order spherical harmonics (SH) model for color estimation. Specifically, the RGB value decoded from the semantic features is directly used as the Gaussian color, which keeps the appearance model simple and efficient. Nevertheless, both the decoded depth and color are only coarse approximations, and relying on them alone may lead to noticeable degradation in rendering quality.

\textbf{Color and Depth Residual.}
\textbf{Color and Depth Residual.}
Accurate color and depth are crucial for high-fidelity rendering. To further refine these two attributes, we estimate per-view residuals during Gaussian-parameter prediction:
\begin{align}
\mathbf{C}_{r,t}^V &= \mathrm{Tanh}\left(h_c\left(f_p\left(\mathcal{T}(\hat{\mathbf{f}}_t^V), \hat{\mathbf{F}}_t^V\right)\right)\right), \nonumber  \\
\mathbf{Z}_{r,t}^V &= \mathrm{Tanh}\left(h_d\left(f_p\left(\mathcal{T}(\hat{\mathbf{f}}_t^V), \hat{\mathbf{F}}_t^V\right)\right)\right),
\end{align}
\noindent where $\mathbf{C}_{r,t}^V$ and $\mathbf{Z}_{r,t}^V$ denote the residuals for color and depth, respectively. The fusion function $f_p(\cdot)$ integrates local and global contextual information, and the Tanh activation bounds the residual magnitude.

The refined color and depth are then obtained as
\begin{align}
\mathbf{C}_t^V &= \hat{\mathbf{X}}_t^V + \mathbf{C}_{r,t}^V, \nonumber  \\
\mathbf{Z}_t^V &= \mathcal{T}(\hat{\mathbf{D}}_t^V) + \mathbf{Z}_{r,t}^V,
\end{align}
\noindent where $\hat{\mathbf{X}}_t^V$ and $\hat{\mathbf{D}}_t^V$ are the decoded color and disparity, respectively. This refinement step alleviates the limitations of the coarse estimates and compression artifacts, thereby enhancing the overall rendering quality. The refined depth maps are subsequently back-projected from pixel coordinates into 3D space to determine the center position of each Gaussian in a pixel-wise manner:
\begin{align}
\boldsymbol{\mu}_t^V &= \Pi^{-1}_{\mathbf{p}_t^V}( \mathbf{Z}_t^V ),
\end{align}
where $\Pi^{-1}_{\mathbf{p}_t^V}(\cdot)$ denotes the inverse projection given pixel coordinates $\mathbf{p}_t^V$. 
Given the camera parameters for each view, we construct the projection matrix $\mathbf{P}^V$ to lift Gaussians defined on the image plane into 3D space. By directly deriving these 3D attributes from the shared semantic features, our design inherently avoids the repeated, high-complexity extraction of inter-view correlations, thereby fundamentally resolving the computational redundancy flaw of decoupled paradigms identified in Section~I.

\subsection{Training Strategy}

We train the entire architecture in an end-to-end manner, including disparity estimation and compression, joint multi-view video coding, Gaussian parameter prediction, and differentiable rendering. 
At each time step $t$, the overall objective follows a RRD formulation:
\begin{equation}
\mathcal{L}_t = \lambda_t D_t + \gamma \mathcal{L}^d_t + R_t,
\end{equation}
where $R_t$ denotes the bitrate for compressing the left and right frames into latent features, and $\lambda_t$ is a time-dependent trade-off weight. 

The term $\mathcal{L}^d_t$ is the disparity estimation loss~\cite{lipson2021raft} with a fixed weighting factor $\gamma = 0.05$. Since our framework processes stereo views in parallel, this loss is computed over both branches:
\begin{equation}
\mathcal{L}^d_t = \sum_{V \in \{L,R\}} \sum_{k=1}^{K} \mu^{K-k} \left\| \mathbf{D}_{t,\text{gt}}^V - \mathbf{D}_t^{V,k} \right\|_1,
\end{equation}
where $\mathbf{D}_{t,\text{gt}}^V$ is the ground-truth disparity and $\{\mathbf{D}_t^{V,k}\}_{k=1}^{K}$ are the iteratively refined disparity estimates for view $V$.

The overall distortion term $D_t$ jointly accounts for novel-view rendering quality and multi-view source reconstruction:
\begin{equation}
D_t = \alpha \mathcal{L}^{SSIM}_t + (1 - \alpha) \mathcal{L}^1_t + \beta \mathcal{L}^{src}_t.
\end{equation}
Here, the rendering distortion ($\mathcal{L}^{SSIM}_t$ and $\mathcal{L}^1_t$) is measured between the synthesized novel views and the corresponding ground-truth novel views with $\alpha = 0.2$, following~\cite{kerbl20233d}. 
In addition, we introduce a reconstruction loss on the multi-view source frames, weighted by $\beta = 0.2$:
\begin{equation}
\mathcal{L}^{src}_t = \left\| \mathbf{X}^L_t - \hat{\mathbf{X}}^L_t \right\|_2^2 + \left\| \mathbf{X}^R_t - \hat{\mathbf{X}}^R_t \right\|_2^2,
\end{equation}
where $\|\cdot\|_2^2$ denotes the mean squared error (MSE), while $\mathbf{X}_t^V$ and $\hat{\mathbf{X}}_t^V$ are the original and reconstructed stereo frames, respectively.

For a video sequence of length $T$, the total training objective is given by:
\begin{equation}
\mathcal{L} = \sum_{t=1}^T \big( \lambda_t D_t + R_t + \gamma \mathcal{L}^d_t \big).
\end{equation}

Hierarchical quality structures have been shown to improve average compression performance~\cite{DCVCDC,DCVCFM}. 
Leveraging this insight, we employ periodically varying weights $\lambda_t$ across frames, as described in our QP scheduling strategy, which further enhances the end-to-end compression efficiency and rendering quality.

\section{Experiments}
\subsection{Settings}

\textbf{Datasets.} We evaluate our method on three dynamic human datasets: two scanned datasets (4D-DRESS~\cite{wang20244d} and X-Humans~\cite{shen2023x}) and one real-world dataset (DNA-Rendering~\cite{cheng2023dna}). 4D-DRESS serves as our primary training and validation set. For each 30-frame sequence, we select adjacent camera pairs from a 16-camera circular array as stereo inputs, and sample intermediate viewpoints for ground-truth rendering. To rigorously validate cross-domain generalization, we directly test on X-Humans and the real-world DNA-Rendering dataset using the identical stereo sampling and rendering protocol, without any additional fine-tuning.

\textbf{Implementation details.} 
All experiments are conducted with GS-SCNet implemented in PyTorch and trained on a single NVIDIA A100 GPU using the AdamW optimizer. 
Training follows two successive stages. First, only the disparity-estimation branch and the compression sub-networks are optimized using a single rate--distortion trade-off weight. Second, the entire pipeline, including rendering, is fine-tuned end-to-end. The learning rate is initialized to $1 \times 10^{-4}$ and the fine-tuning phase runs for 150k iterations. 
The codec supports 64 different quantization parameters (QPs) mapping to $\lambda \in [1, 750]$ following~\cite{DCVCFM}. We instantiate models at six representative QP points $\{7, 15, 23, 31, 39, 47\}$. Additionally, we introduce a temporal layering strategy where the nominal QP is periodically perturbed by an offset pattern $\{0, 8, 0, 4\}$ to induce quality scalability along the temporal dimension.

\textbf{Evaluation metrics.} 
Bitrate is reported in bits per pixel. Since the final output is a rendered novel-view sequence, distortion is directly measured between rendered views and ground-truth images using PSNR, SSIM, and Learned Perceptual Image Patch Similarity (LPIPS) to evaluate perceptual quality. We refer to this formulation as a \emph{rate--rendering distortion} problem. To compare overall efficiency, we additionally compute the Bjøntegaard-Delta rate (BD-rate)~\cite{bjontegaard2001calculation} under PSNR and SSIM.

\textbf{Baselines.} 
For a comprehensive comparison against our end-to-end GS-SCNet, we construct decoupled baselines by combining established video codecs with state-of-the-art feed-forward 3DGS networks. Specifically, we independently compress the multi-view RGB streams using H.265/HEVC~\cite{H265}, MV-HEVC~\cite{MVHEVC}, FVC~\cite{hu2021fvc}, DCVC-DC~\cite{DCVCDC}, and DCVC-RT~\cite{DCVCRT}. The decoded videos are then fed into either GPS-Gaussian~\cite{zheng2024gps} or EVA-Gaussian~\cite{hu2024eva} for novel view synthesis. Both NVS backends are retrained on 4D-DRESS under identical settings to ensure a fair RRD evaluation.

\subsection{Comparison Results}

\begin{figure*}[t]
    \centering
    \includegraphics[width=0.45\textwidth]{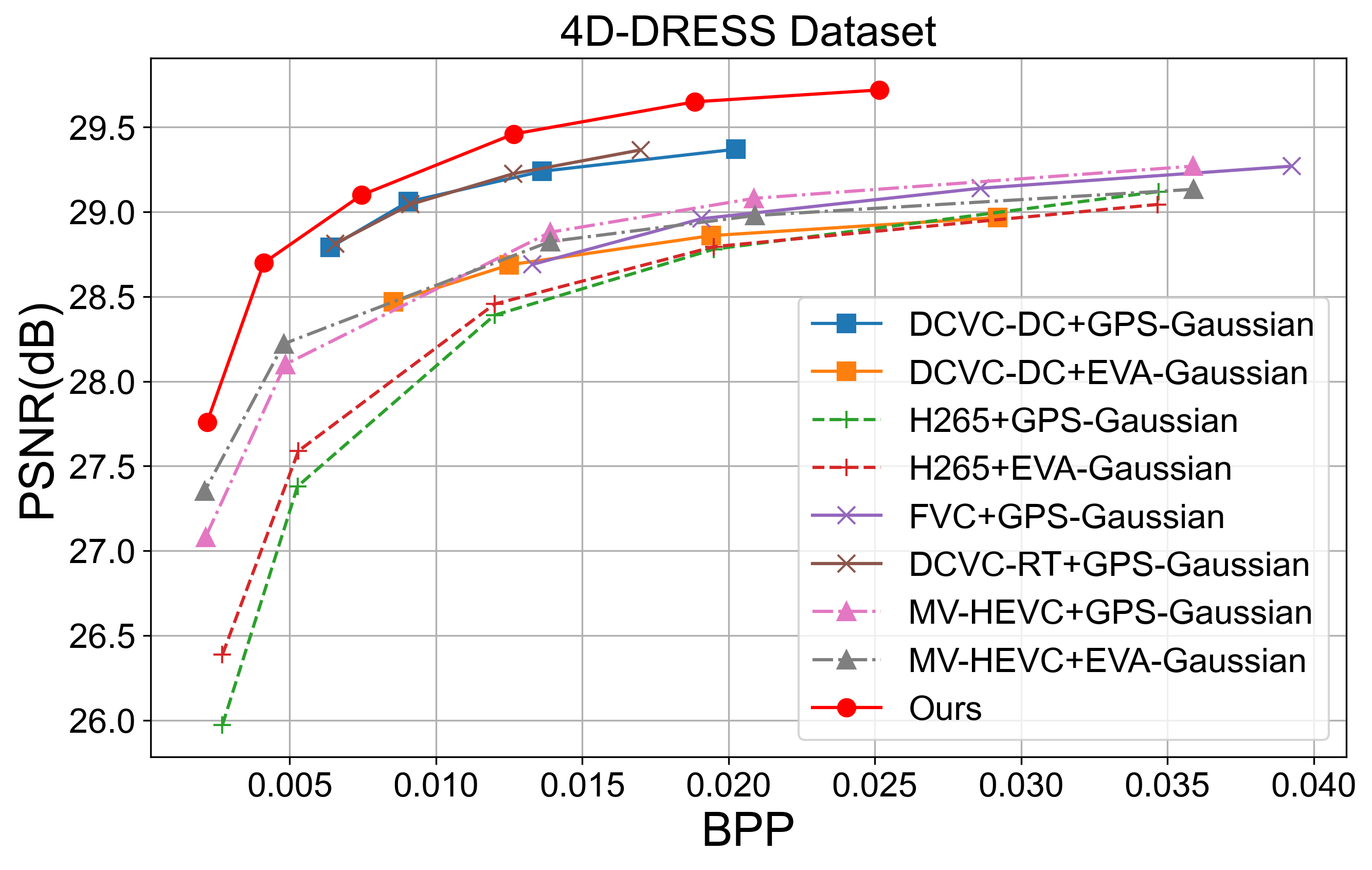}
    \includegraphics[width=0.45\textwidth]{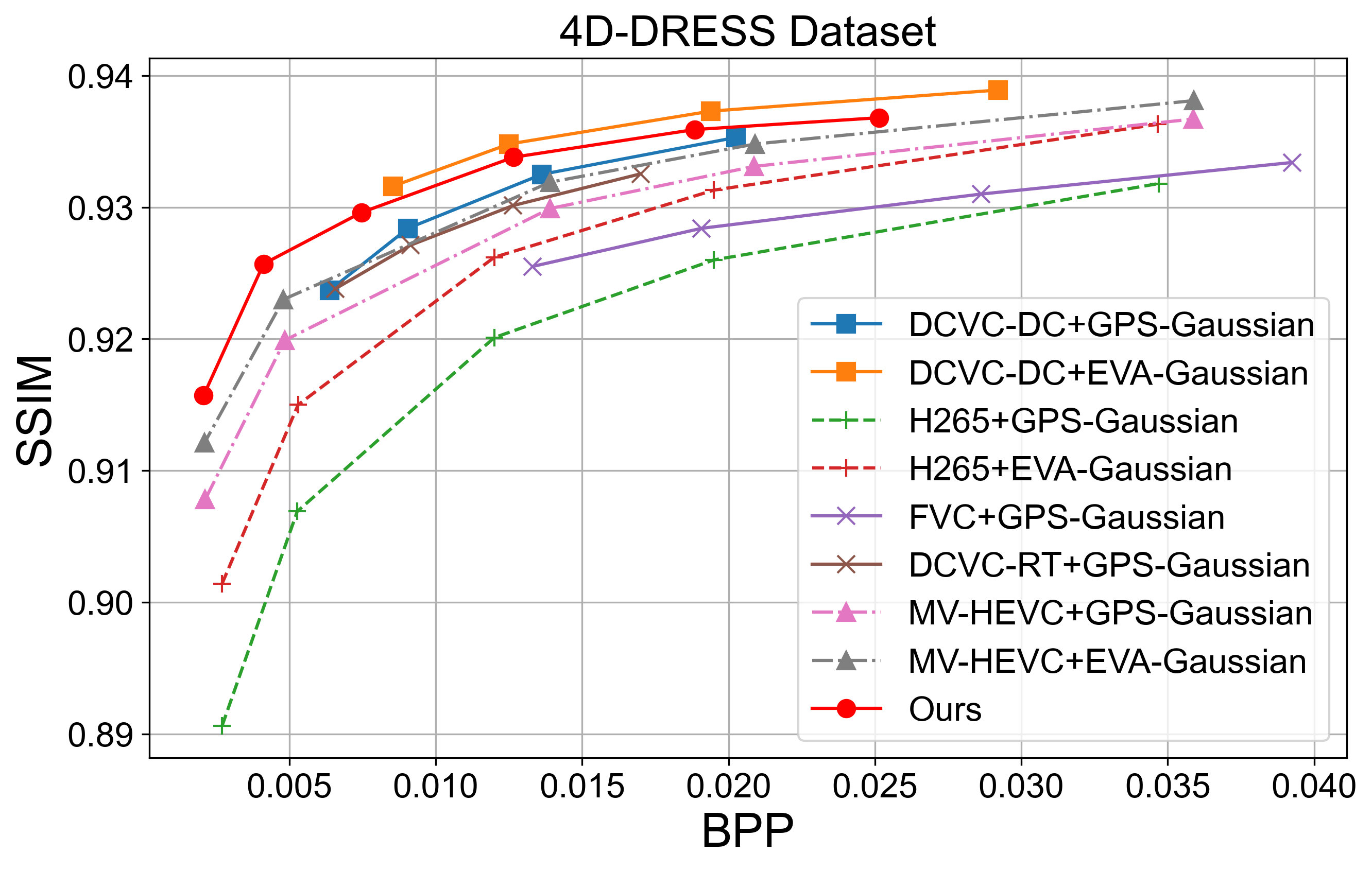} \\
    \includegraphics[width=0.45\textwidth]{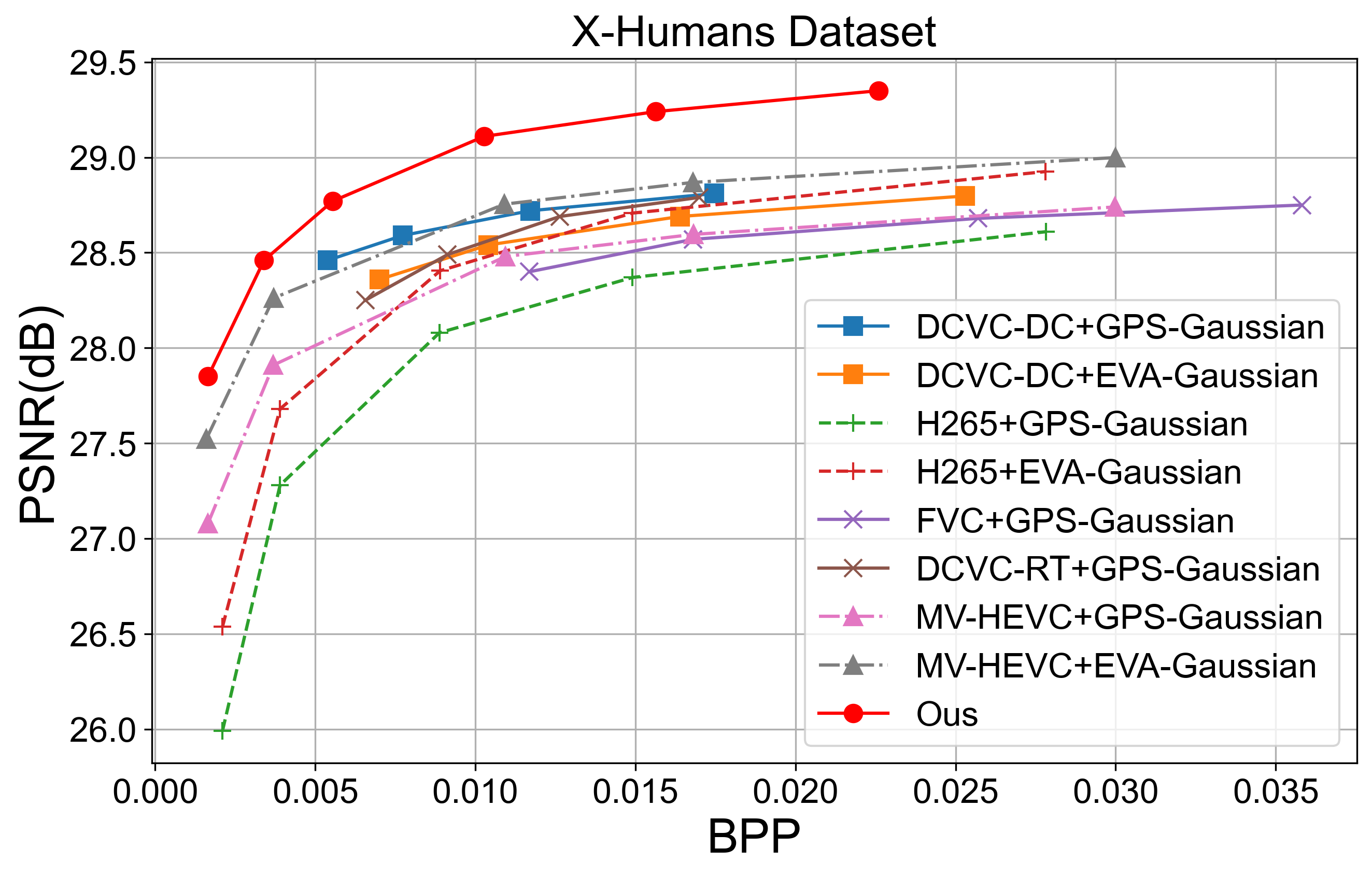}
    \includegraphics[width=0.45\textwidth]{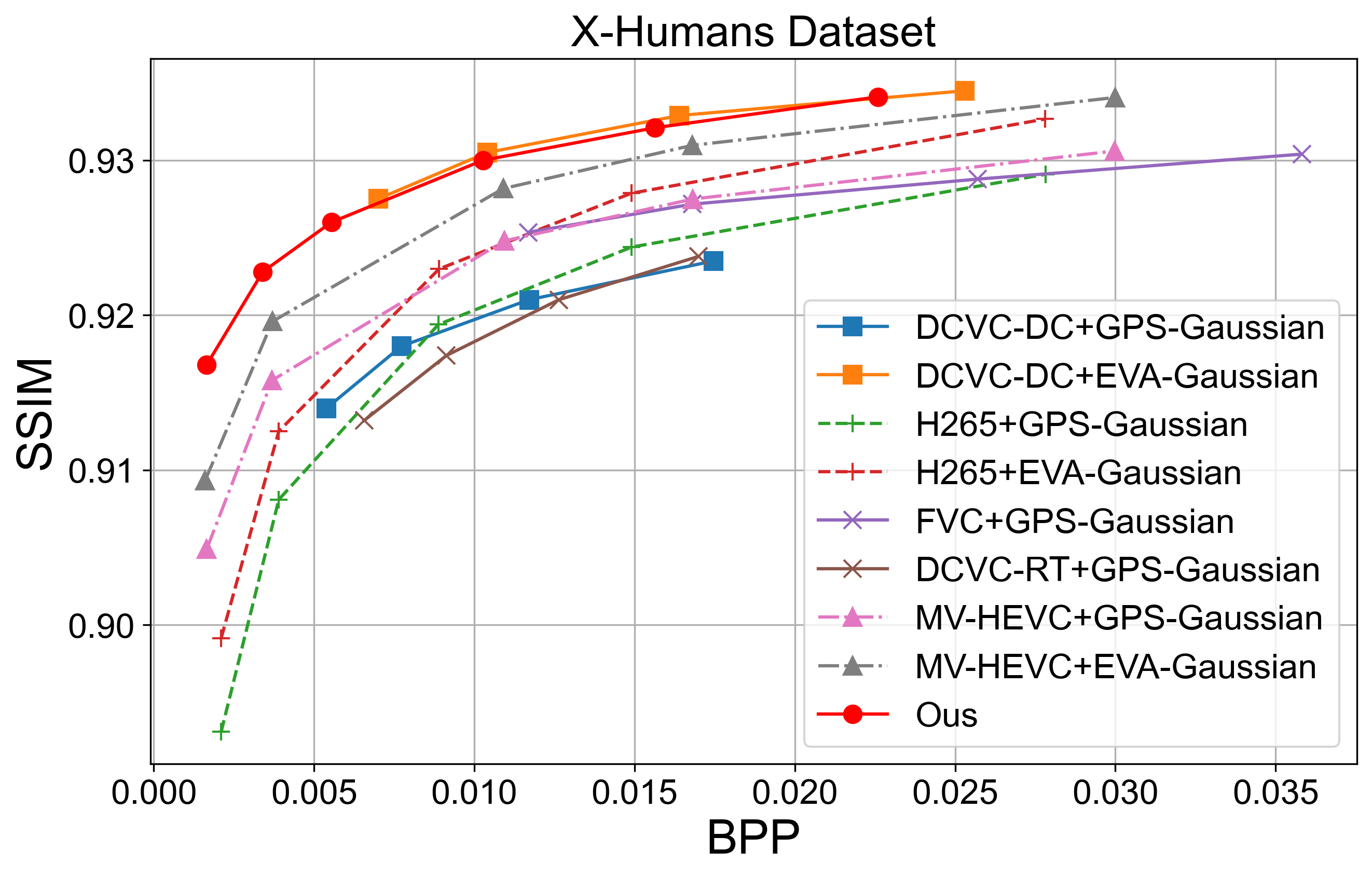}
    \caption{RRD curves in terms of PSNR and SSIM on the 4D-DRESS~\cite{wang20244d} and X-Humans~\cite{shen2023x} datasets.}
    \label{fig:rd_metrics}
\end{figure*}

\begin{table}[t]
\centering
\caption{BD-Rate(\%) comparison measured with PSNR, using MV-HEVC+GPS-Gaussian as the anchor.}
\label{tab:bd-rate1}
\begin{tabular}{lccc}
\toprule
 & 4D-DRESS & X-humans  & Average  \\
\midrule
MV-HEVC+GPS-Gaussian & 0.0    & 0.0     & 0.0 \\
H265+GPS-Gaussian    & 80.38  & 83.40   & 81.89\\
FVC+GPS-Gaussian     & 19.31  & 15.43   & 17.37\\
DCVC-DC+GPS-Gaussian & -53.81 & -52.85  & -53.33\\
DCVC-RT+GPS-Gaussian & -53.50 & -23.65  & -38.58\\
H265+EVA-Gaussian    & 67.40  & 58.02   & 62.71\\
MV-HEVC+EVA-Gaussian & -5.76  & -42.99  & 24.38\\
DCVC-DC+EVA-Gaussian & 35.47 & -37.75  & -1.14\\
\textbf{GS-SCNet}    & -67.46 & -82.89  & -75.18\\

\bottomrule
\end{tabular}
\end{table}

\begin{table}[t]
\centering
\caption{BD-Rate(\%) comparison measured with SSIM, using MV-HEVC+GPS-Gaussian as the anchor.}
\label{tab:bd-rate2}
\begin{tabular}{lccc}
\toprule
 & 4D-DRESS & X-humans  & Average  \\
\midrule
MV-HEVC+GPS-Gaussian & 0.0    & 0.0     & 0.0 \\
H265+GPS-Gaussian    & 137.39 & 67.21   & 102.30\\
FVC+GPS-Gaussian     & 71.01  & 11.81   & 41.41\\
DCVC-DC+GPS-Gaussian & -22.72 & -25.46  & -24.09\\
DCVC-RT+GPS-Gaussian & -9.15 & -11.59  & -10.37\\
H265+EVA-Gaussian    & 40.49  & 28.54   & 34.52\\
MV-HEVC+EVA-Gaussian & -24.09  & -34.48   & -29.29\\
DCVC-DC+EVA-Gaussian & -52.45 & -75.64  & -64.05\\
\textbf{GS-SCNet}    & -44.56 & -60.74  & -52.65\\
\bottomrule
\end{tabular}
\end{table}

\textbf{Quantitative RRD Performance.} Fig.~\ref{fig:rd_metrics} plots the RRD curves of our method on the 4D-DRESS and X-Humans datasets, where we compare the end-to-end compression and NVS performance against decoupled pipelines. These baselines apply conventional or learned video codecs followed by advanced feed-forward human 3DGS reconstruction methods (GPS-Gaussian~\cite{zheng2024gps} and EVA-Gaussian~\cite{hu2024eva}). On both datasets, GS-SCNet consistently outperforms these decoupled coding schemes. This significant superiority fundamentally stems from two architectural advantages. First, unlike decoupled learned single-view codecs (e.g., DCVC-DC, FVC) that leave multi-view redundancy uneliminated, our disparity-guided semantic coding efficiently compresses cross-view correlations, leading to substantial bitrate savings. Second, decoupled codecs strictly optimize for RD, blindly allocating bits to high-frequency pixel noises that do not contribute to 3D geometry. In contrast, by directly optimizing the Rate-Rendering Distortion objective, our end-to-end method intelligently allocates bitrates to semantic features that are most crucial for rendering quality. Furthermore, our model delivers highly competitive and robust performance on X-Humans without any fine-tuning, indicating strong generalization across diverse human subjects. It is worth noting that although the decoupled pipeline of DCVC-DC followed by EVA-Gaussian achieves the best performance in terms of SSIM, its reliance on direct depth estimation leads to poor robustness when applied to data with depth distributions significantly different from the training set, such as the real-world DNA-Rendering dataset (which will be analyzed in detail in subsequent sections).

Table~\ref{tab:bd-rate1} and Table~\ref{tab:bd-rate2} report the BD-Rate comparisons under PSNR and SSIM, respectively. Taking MV-HEVC+GPS-Gaussian as the reference, our end-to-end framework achieves average BD-Rate reductions of 75.1\% in terms of PSNR and 51.66\% in terms of SSIM over the two datasets. Even when compared with recent learned codecs, GS-SCNet maintains a clear margin: relative to DCVC-DC+GPS-Gaussian, our approach yields BD-Rate savings of 35.64\% (PSNR) and 42.62\% (SSIM); when DCVC-RT+GPS-Gaussian is used as the baseline, GS-SCNet further improves the average BD-Rate by 44.64\% and 21.57\% for PSNR and SSIM, respectively. These comprehensive results demonstrate that GS-SCNet not only compresses multi-view video streams more efficiently than both conventional and advanced learned codecs, but also offers a significantly more robust and higher-quality novel view synthesis.

\begin{figure}[t]
  \centering
  \includegraphics[width=\linewidth]{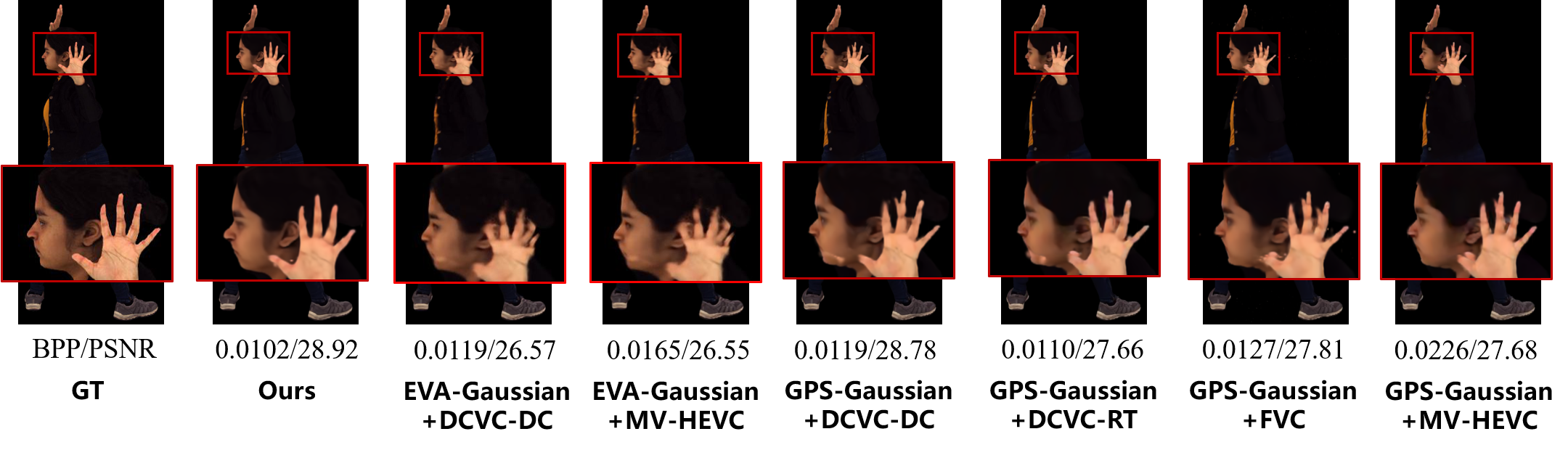}
  \caption{Rendering quality comparison on 4D-DRESS dataset.}
  \label{quality_comparision}
\end{figure}

\begin{figure}[t]
  \centering
  \includegraphics[width=\linewidth]{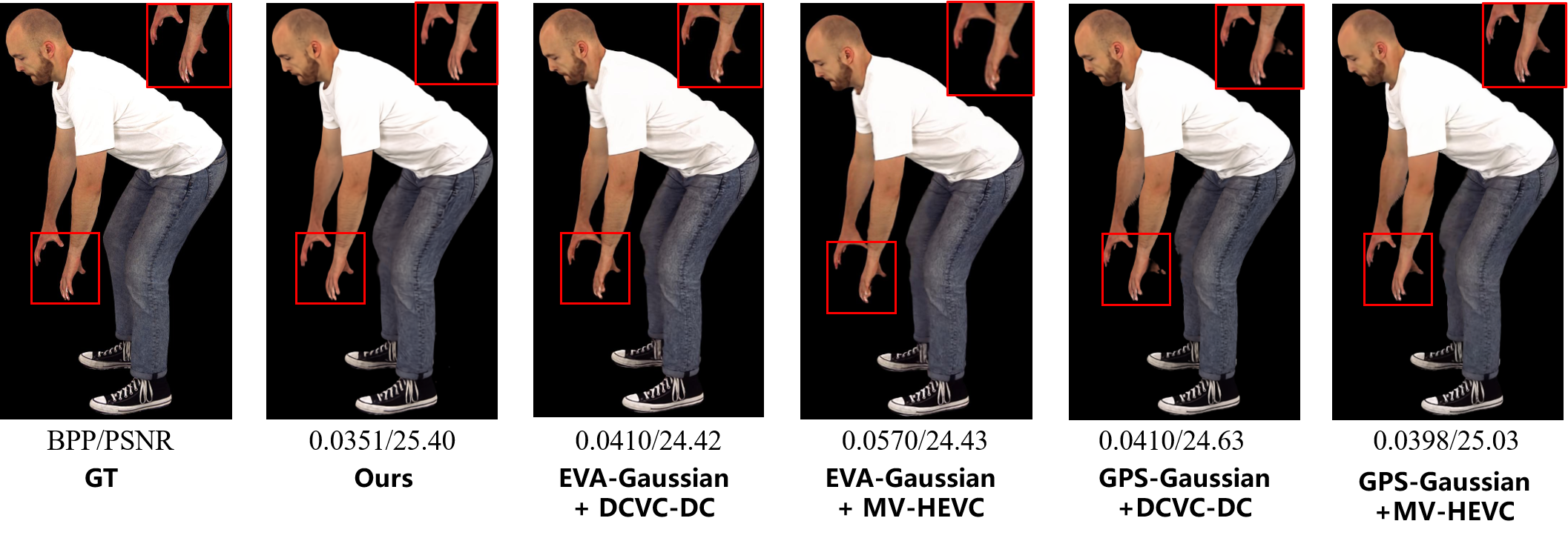}
  \caption{Rendering quality comparison on X-Humans dataset.}
  \label{quality_comparision_x}
\end{figure}


\textbf{Rendering Quality.} 
Fig.~\ref{quality_comparision} and Fig.~\ref{quality_comparision_x} show qualitative comparisons of rendering results on the 4D-DRESS and X-Humans datasets, respectively. Given the extensive number of evaluated baseline combinations, we display representative selections to maintain visual clarity; specifically, Fig.~\ref{quality_comparision} provides a comprehensive comparison across the decoupled pipelines, whereas Fig.~\ref{quality_comparision_x} focuses on four representative baselines for conciseness. In the decoupled paradigms, conventional codecs (e.g., H.265, MV-HEVC) tend to introduce visible blocking artifacts and blur fine details. Meanwhile, even when employing advanced learning-based video codecs (e.g., FVC, DCVC-DC, DCVC-RT) alongside state-of-the-art NVS backends like GPS-Gaussian and EVA-Gaussian, these decoupled methods still struggle to faithfully preserve subtle structures. For instance, as clearly observed in the zoomed-in patches of both datasets, the decoupled baselines suffer from severe blurring, geometric distortion, and fusion artifacts when rendering complex hand gestures and facial contours. In stark contrast, under similar or even lower bitrates, GS-SCNet produces significantly sharper contours, clearer facial features, and more precise finger structures, all of which are crucial for immersive communication. These observations visually confirm the superiority of our disparity-guided semantic compression and end-to-end lightweight Gaussian-parameter prediction design.

\begin{figure}[t]
  \centering
  \includegraphics[width=\linewidth]{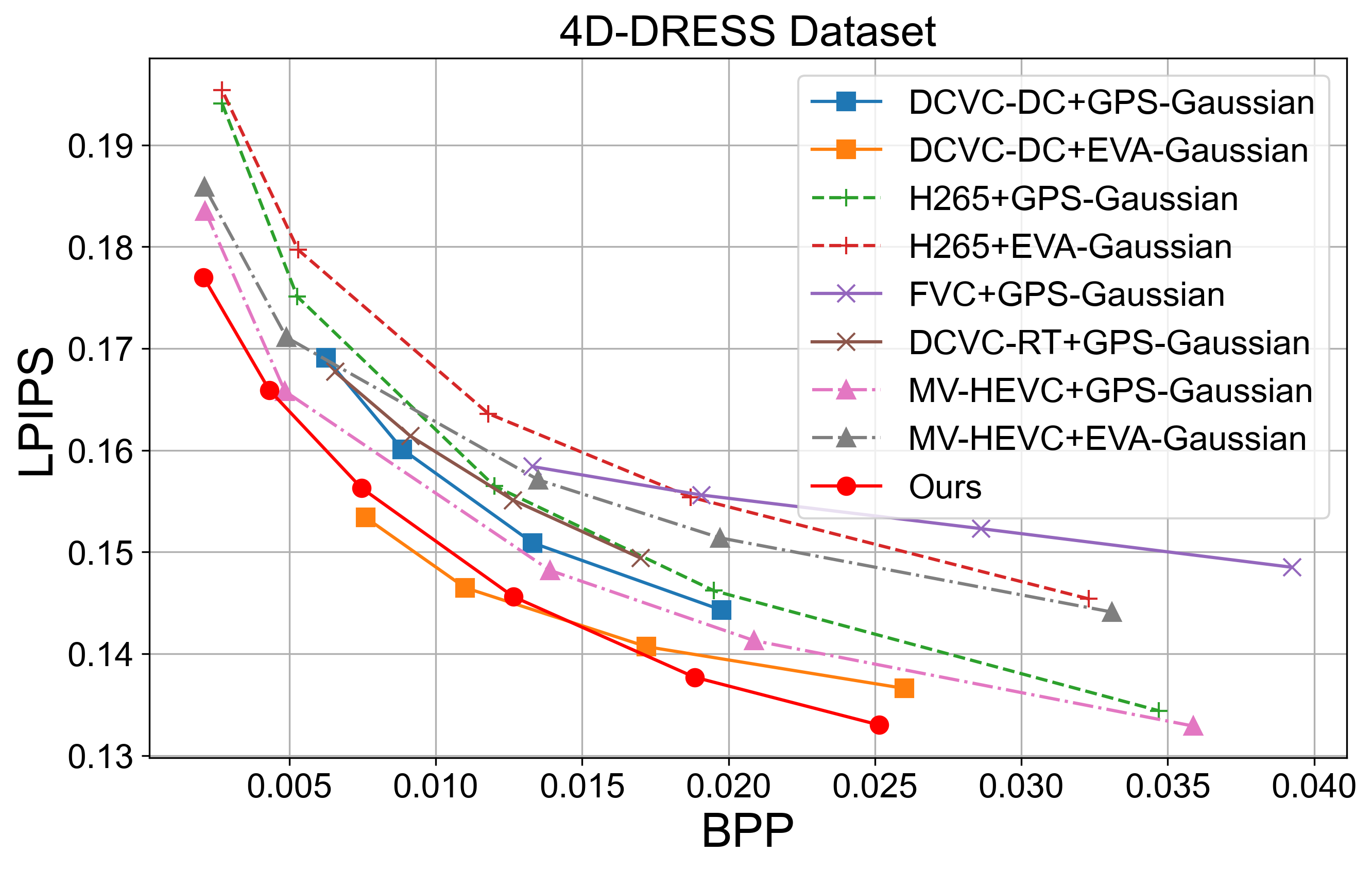}
  \caption{RRD curves in terms of LPIPS on the 4D-DRESS dataset.}
  \label{LPIPS}
\end{figure}

\textbf{Perceptual Quality.} Beyond PSNR and SSIM, Fig.~\ref{LPIPS} reports RD curves with LPIPS on the 4D-DRESS dataset. GS-SCNet consistently yields lower LPIPS values across all bitrate ranges compared to other baselines, confirming its advantage in perceptual quality. 


\subsection{Robustness and Generalization on Real-World Data}


\begin{figure*}[t]
    \centering
    \includegraphics[width=0.31\textwidth]{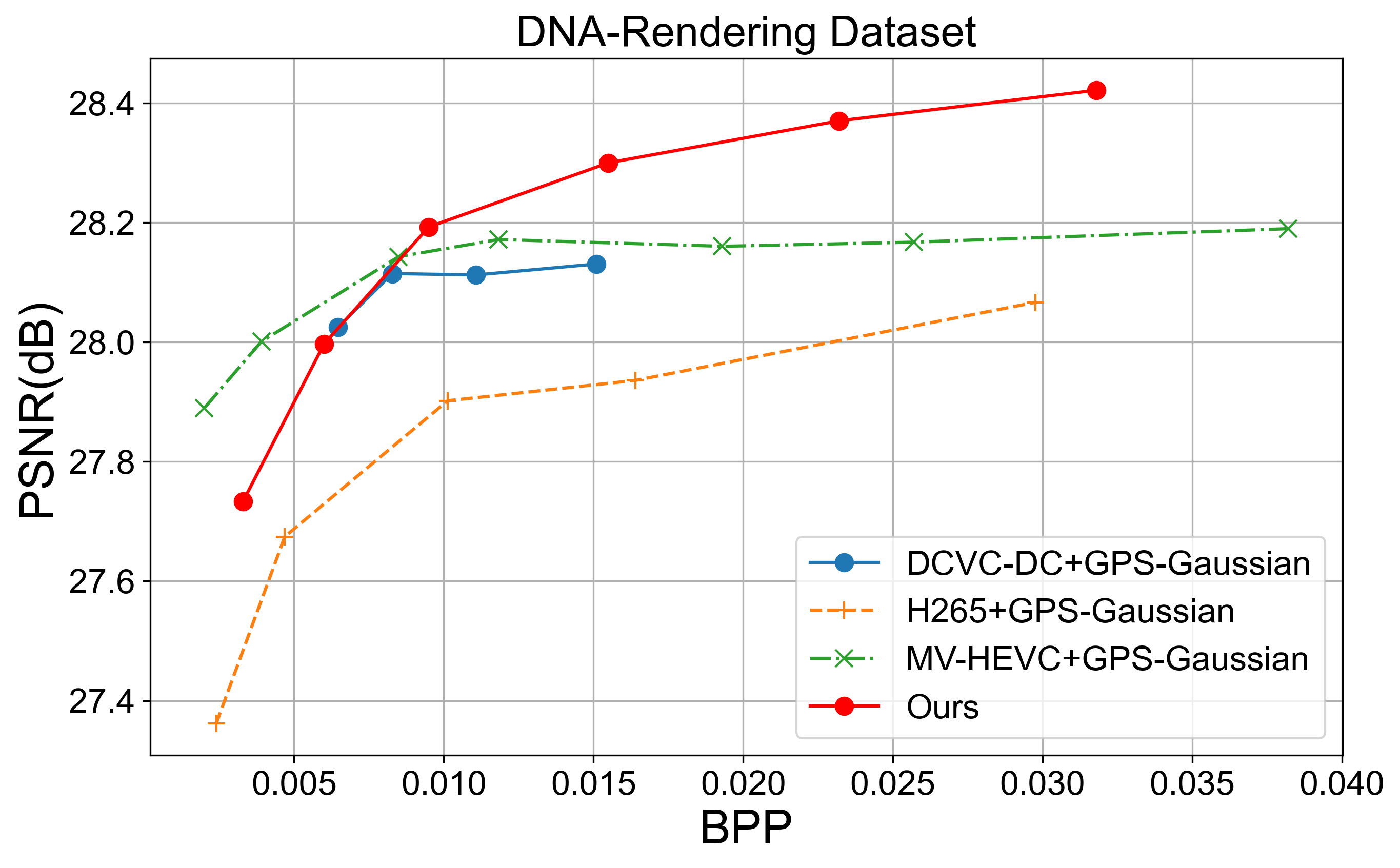}
    \includegraphics[width=0.31\textwidth]{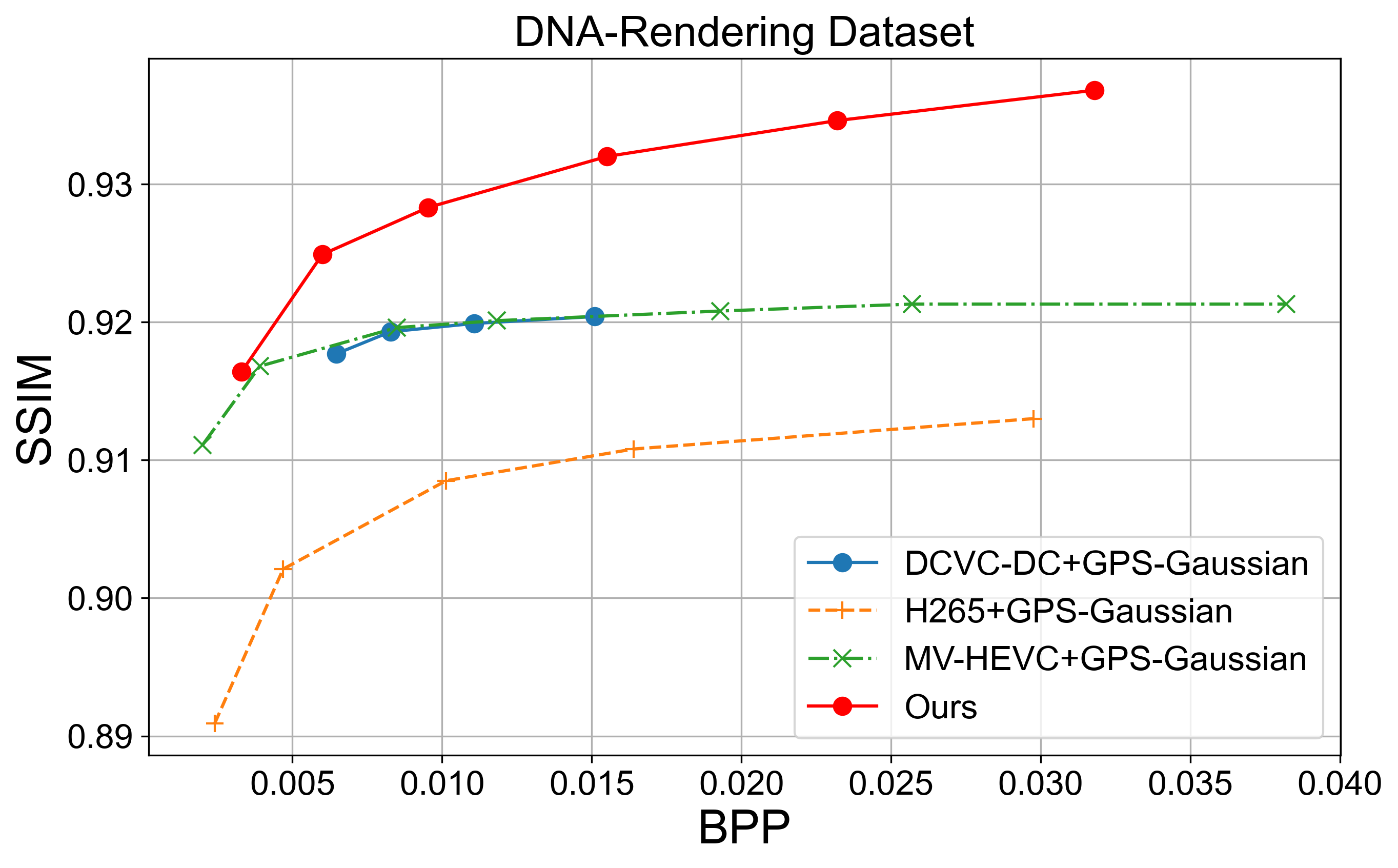} 
    \includegraphics[width=0.31\textwidth]{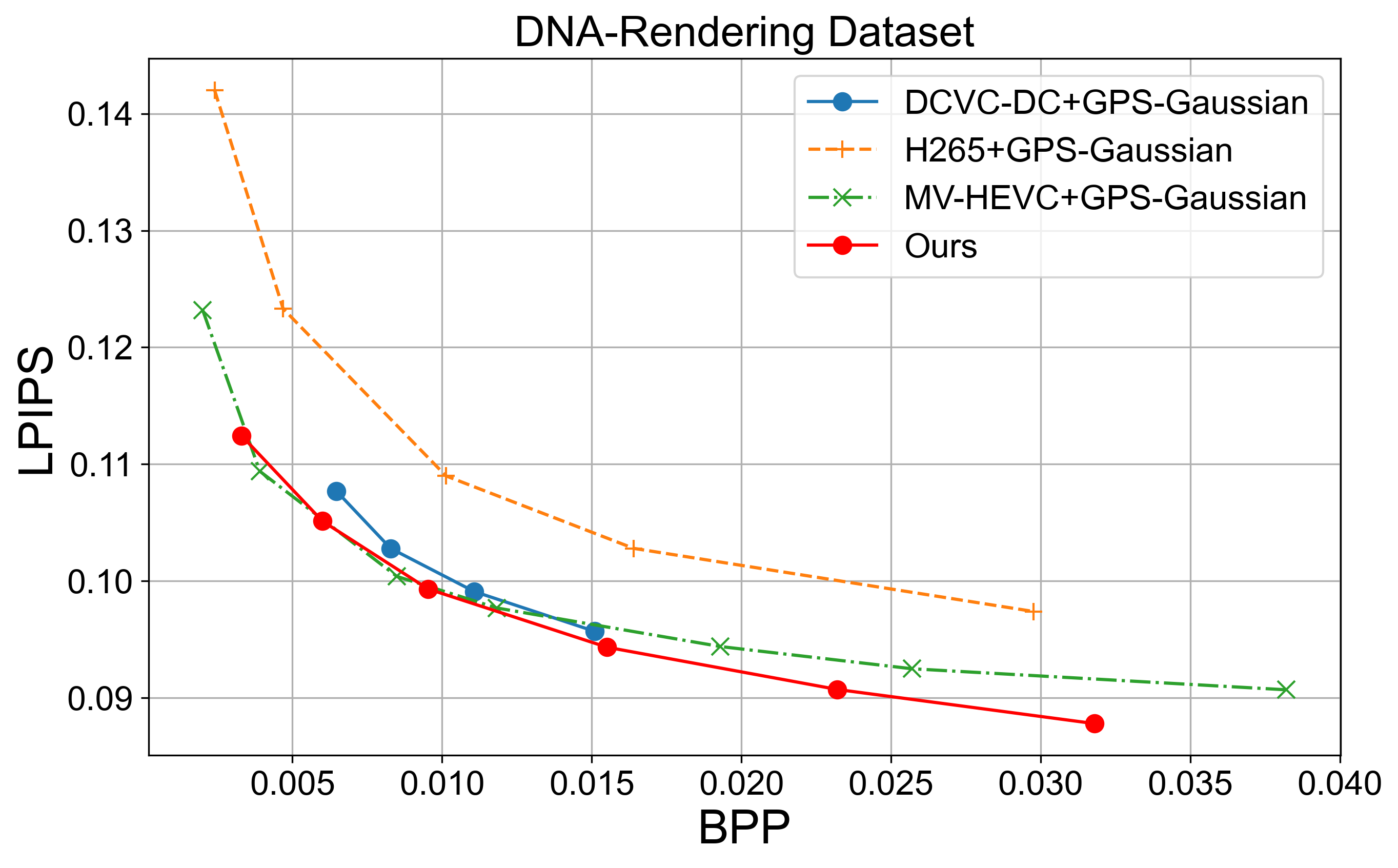}
    \caption{RRD curves in terms of PSNR, SSIM and LPIPS on the DNA-Rendering\cite{cheng2023dna} datasets.}
    \label{DNA}
\end{figure*}

\begin{figure}[th]
  \centering
  \includegraphics[width=\linewidth]{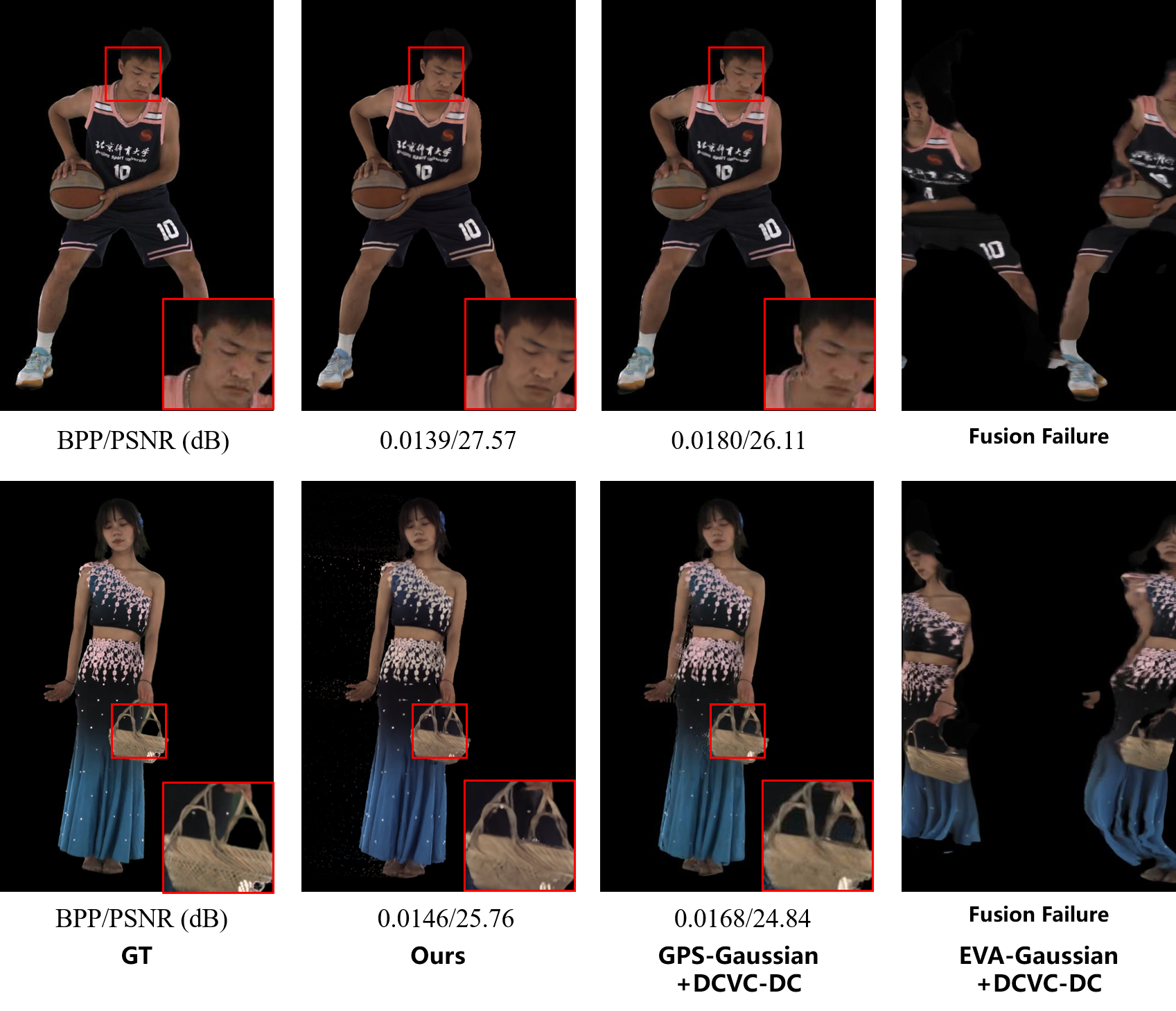}
  \caption{Qualitative comparison of novel view synthesis on the real-world DNA-Rendering dataset. While our GS-SCNet produces accurate geometry and sharp high-frequency details, the decoupled GPS-Gaussian pipeline suffers from blurriness. Notably, the decoupled EVA-Gaussian pipeline struggles with scale ambiguity, resulting in noticeable structural misalignment and ghosting artifacts.}
  \label{quality}
\end{figure}

To robustly validate the generalization capability of our framework, we conduct cross-domain evaluations on the real-world DNA-Rendering dataset using models solely trained on synthetic data, without any fine-tuning. We select three adjacent cameras positioned at the same height: the two outer cameras provide the stereo inputs, while the middle one serves as the target novel view. As illustrated by the RRD curves in Fig.~\ref{DNA}, GS-SCNet degrades gracefully and significantly outperforms the decoupled GPS-Gaussian pipeline across all bitrates. We deliberately omit the quantitative curve for the decoupled EVA-Gaussian pipeline due to its severe depth estimation errors on this out-of-domain dataset, which lead to catastrophic structural ghosting. Instead, we visualize its reconstruction breakdown in Fig.~\ref{quality}.

As observed in Fig.~\ref{quality}, GS-SCNet successfully preserves high-frequency features (e.g., facial details and textures) with virtually no compression artifacts under similar bitrates. While GPS-Gaussian correctly estimates the overall geometry, it suffers from noticeable blurriness. In stark contrast, EVA-Gaussian fails entirely to align and fuse the 3DGS representations extracted from the two input sources, resulting in a severe "fusion failure" artifact.

The root cause of this catastrophic failure lies in the fundamental difference in depth estimation mechanisms. Both our GS-SCNet and GPS-Gaussian estimate relative disparity first, which is deterministically converted into absolute depth using the known camera baseline. By explicitly leveraging ground-truth camera parameters, this mechanism circumvents the scale ambiguity ubiquitous in monocular 3D reconstruction, correctly predicting the subject-to-camera distance to be approximately 3 meters on the DNA-Rendering dataset. Conversely, EVA-Gaussian directly predicts absolute depth from the input images, making it highly susceptible to domain overfitting. Since it is primarily trained on synthetic datasets (e.g., THuman2.0) where virtual cameras are fixed at roughly 2 meters, the pre-trained model stubbornly applies this 2-meter absolute scale to the real-world data. Consequently, during the unprojection process, the 3D Gaussians from the two views are projected into disjointed 3D spaces, causing the observed structural split.

In summary, by explicitly utilizing camera parameters to eliminate scale ambiguity, GS-SCNet achieves significantly more robust geometric estimation. Benefiting from this robust geometry, our end-to-end framework reconstructs high-fidelity human representations at comparable bitrates, vastly outperforming decoupled transmission paradigms in real-world, out-of-domain scenarios.

\subsection{Ablation Studies}

To assess the contribution of each module, we perform an extensive ablation study in which all model variants are trained and evaluated on the 4D-DRESS dataset under a unified setting.


\begin{figure*}[htbp]
    \centering
    \subfloat[Cross-view parallel autoencoders]{%
        \includegraphics[width=0.45\textwidth]{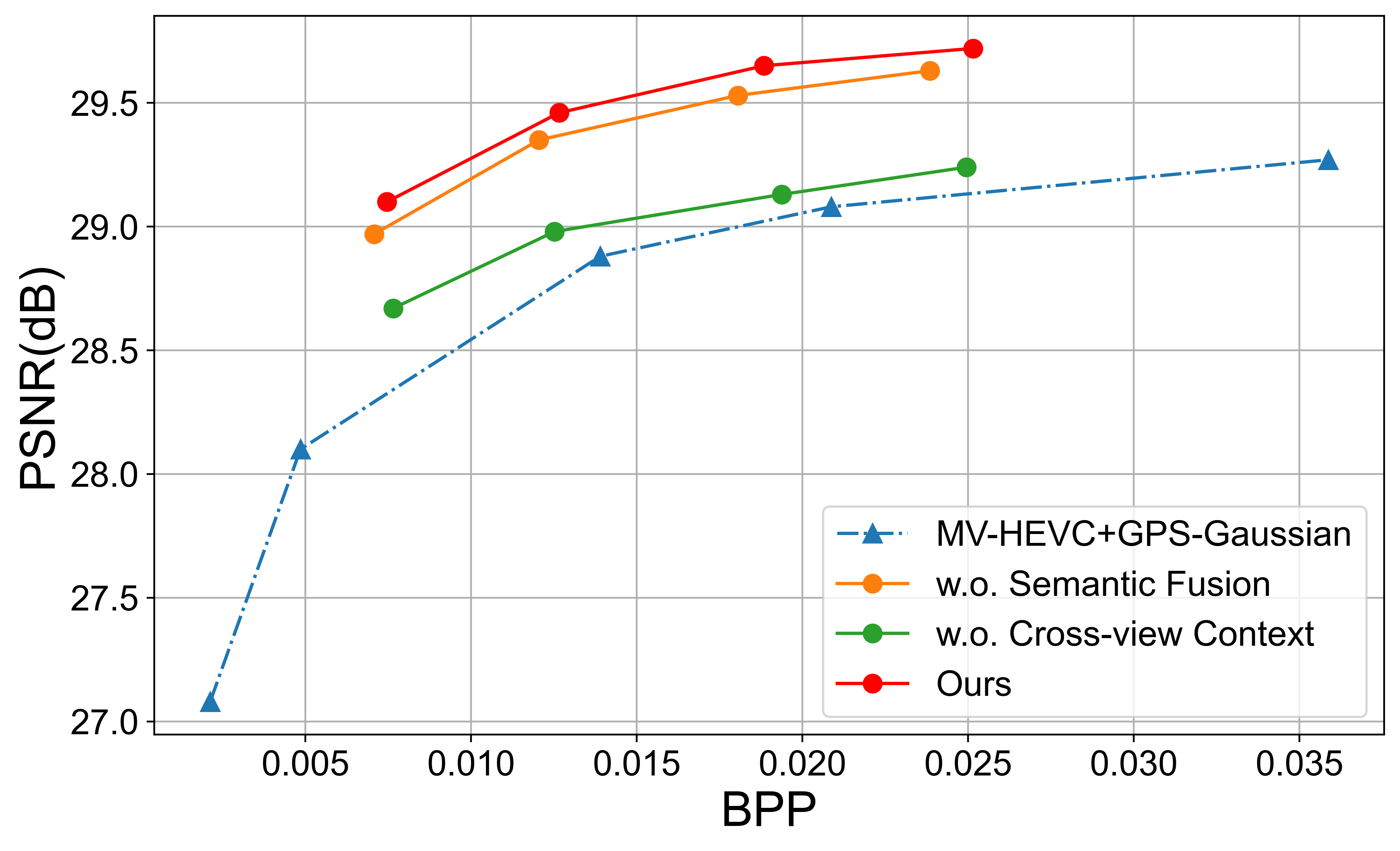}%
        \label{fig:abl1}%
    }\hfill
    \subfloat[Model optimization]{%
        \includegraphics[width=0.45\textwidth]{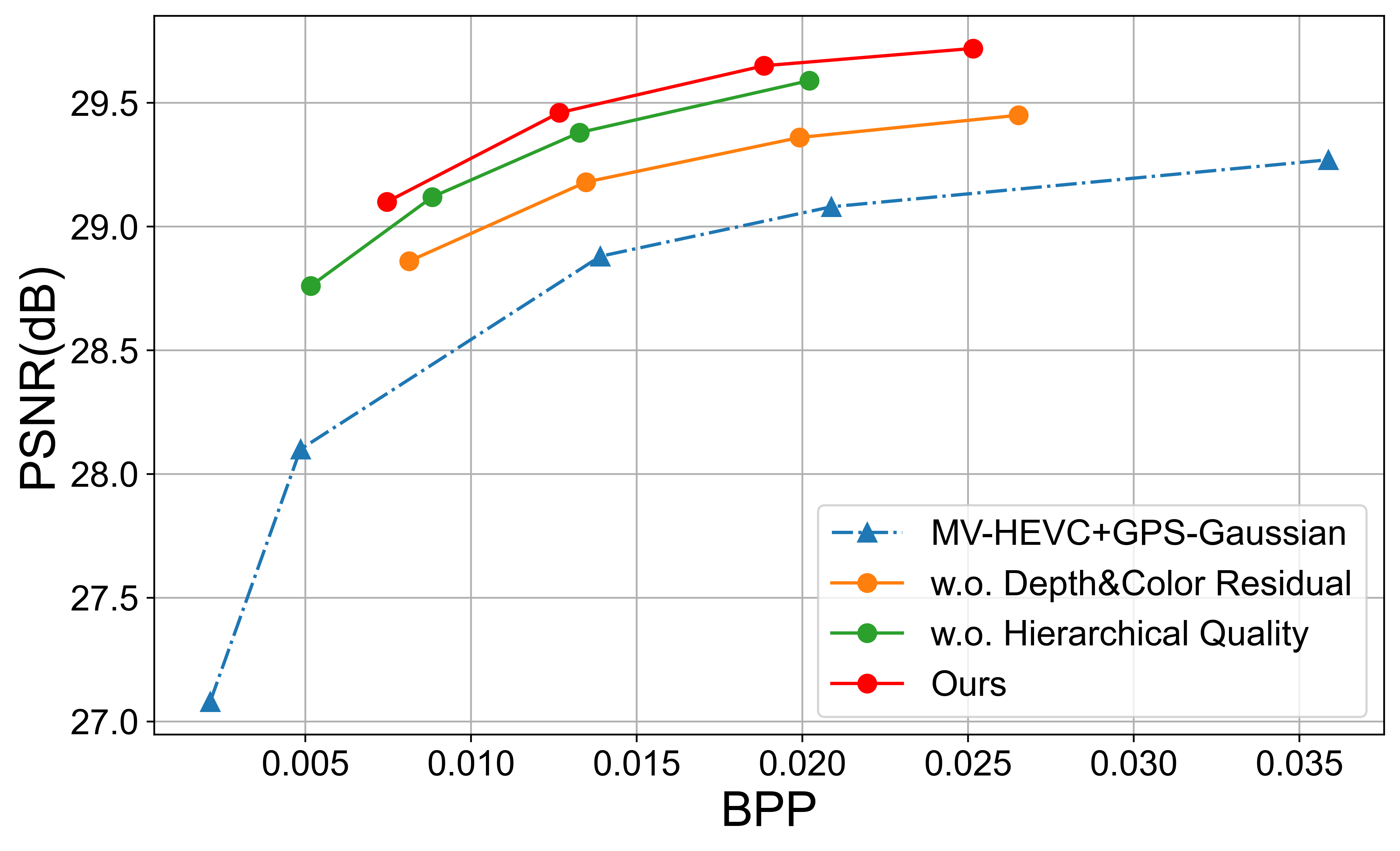}%
        \label{fig:abl2}%
    }%
    \caption{Impact of the cross-view parallel semantic feature autoencoder and model optimization strategies on rate–distortion performance.}
    \label{fig:abl}
\end{figure*}

\begin{table}[t]
\centering
\caption{Ablation study on the 4D-DRESS dataset. The BD-Rate is calculated in terms of PSNR, with our full model ($M_a$) serving as the anchor.}
\label{tab:ablation}
\begin{tabular}{cccccccc}
\toprule
 & \multicolumn{1}{c}{$M_a$} & \multicolumn{1}{c}{$M_b$}  & \multicolumn{1}{c}{$M_c$}  & \multicolumn{1}{c}{$M_d$} & \multicolumn{1}{c}{$M_e$}\\
\midrule
\makecell[c]{Cross-view context \\ w/ semantic fusion}
&\checkmark&          &          &\checkmark&\checkmark&\\
\midrule
\makecell[c]{Cross-view context \\ w/o semantic fusion}
&          &\checkmark&          &          &          & \\
\midrule
Hierarchical quality 
&\checkmark&\checkmark&\checkmark          &&\checkmark&\\
\midrule
Color\&Depth Residual
&\checkmark&\checkmark&\checkmark&\checkmark&          &\\
\midrule
BD-Rate(\%) 
& 0 & 11.86 & 140.62 & 13.61 & 62.15\\
\bottomrule
\end{tabular}
\end{table}

\textbf{Effectiveness of the cross-view parallel semantic feature autoencoder.}  
The cross-view parallel semantic feature autoencoder is a fundamental component of GS-SCNet, responsible for compactly encoding multi-view disparity and image features. To assess its role, we designed ablation settings along two axes: (\emph{i}) removing semantic fusion from the cross-view contextual encoder, and (\emph{ii}) completely removing cross-view context, such that each branch processes its own view independently.  


As shown in Fig.~\ref{fig:abl}(a), eliminating cross-view contextual information ($M_c$) causes a dramatic performance drop, with BD-Rate increasing by 140.62\% compared to the baseline ($M_a$). This highlights the necessity of leveraging inter-view redundancy. Meanwhile, retaining cross-view context but discarding semantic fusion ($M_b$) still degrades performance (+11.86\% BD-Rate). Fundamentally, without semantic fusion, cross-view context is obtained solely via hard disparity-based warping. This direct warping operation is highly vulnerable to occlusions and local geometric mismatches, often indiscriminately transferring noise and artifacts between views. Incorporating the lightweight semantic fusion module mitigates this issue by adaptively evaluating cross-view consistency, effectively suppressing unreliable warping artifacts and producing cleaner, more robust representations for downstream decoding.


\textbf{Effectiveness of model optimization strategies.}  
We also ablate two optimization strategies: (\emph{i}) the hierarchical quality structure, and (\emph{ii}) the color and depth residual modules. Fig.~\ref{fig:abl}(b) and Table~\ref{tab:ablation} show the results. Removing the residual prediction modules ($M_d$) increases BD-Rate by 13.61\%, indicating that directly correcting coarse color and depth predictions is critical for rendering fidelity. Similarly, removing the hierarchical quality structure ($M_e$) causes a large BD-Rate degradation of 62.15\%, demonstrating the importance of quality scalability in optimizing bit allocation.  

Overall, the full model ($M_a$) consistently achieves the best performance. These results confirm that (\emph{i}) cross-view contextual interaction and semantic fusion are indispensable for redundancy reduction, and (\emph{ii}) hierarchical quality modeling and residual refinement provide complementary improvements for compression efficiency and visual quality.

\subsection{Model Complexity and Inference Speed}

\begin{table}[t]
\centering
\caption{COMPUTATIONAL COMPLEXITY AND INFERENCE SPEED. $^*$DENOTES THE SUMMATION OF BOTH SEPARATED MODULES.}
\label{tab:complexity}
\resizebox{\linewidth}{!}{
\begin{tabular}{llccccc}
\toprule
Model & Category & MACs & Params & Tx Time & Rx Time & System FPS \\
\midrule
GPS-Gaussian\cite{zheng2024gps} & NVS & 454G & 4.14M & - & 36ms & - \\
EVA-Gaussian\cite{hu2024eva}    & NVS & 411G & 5.38M & - & 77ms & - \\
\midrule
MV-HEVC + GPS-Gaussian          & Separated & - & - & $\sim$15.9s & 236ms & $< 0.1$ \\
MV-HEVC + EVA-Gaussian          & Separated & - & - & $\sim$15.9s & 277ms & $< 0.1$ \\
DCVC-DC + GPS-Gaussian     & Separated & 1852G$^*$ & 22.66M$^*$ & 173ms & 172ms & 5.8 \\
DCVC-DC + EVA-Gaussian      & Separated & 1809G$^*$ & 23.90M$^*$ & 173ms & 213ms & 4.7 \\
\midrule
\textbf{GS-SCNet (Ours)}        & End-to-End & \textbf{180G} & \textbf{31.36M} & \textbf{52ms} & \textbf{31ms} & \textbf{19.2} \\
\bottomrule
\end{tabular}
}
\end{table}

Table \ref{tab:complexity} evaluates the computational complexity and practical system throughput of our GS-SCNet against existing baselines. While our unified architecture significantly reduces computational overhead (requiring only 180G MACs compared to the severe redundancy in cascaded separated methods), our primary focus lies in real-world deployment speeds. All neural network inference times are strictly measured on an NVIDIA A100 GPU. It is worth noting that the encoding and decoding latencies of MV-HEVC are evaluated using the CPU-based reference software (HTM 16.3), whereas all neural-based codecs and NVS models are accelerated by the GPU. 

To rigorously reflect actual telepresence pipelines, our reported encoding (Tx) and decoding (Rx) latencies implicitly include actual arithmetic entropy coding operations. Crucially, for decoupled paradigms (e.g., DCVC-DC + GPS-Gaussian), the receiver (Rx) latency must accumulate the pixel-domain video decoding time, 3DGS parameter prediction, and final rasterization rendering. Furthermore, assuming a pipelined communication system where the transmitter and receiver operate concurrently, the overall system frame rate (FPS) is bounded by the bottleneck stage ($\frac{1000}{\max(\text{Tx}, \text{Rx})}$). As shown, decoupled methods suffer from severe Rx bottlenecks (e.g., 213 ms for DCVC-DC + EVA-Gaussian) due to the redundant intermediate pixel-domain reconstruction. Even when compared against highly optimized GPU-accelerated decoupled neural codecs like DCVC-DC, our GS-SCNet still achieves a remarkable reduction in inference time. By directly decoding semantic Gaussian parameters in the latent space, GS-SCNet successfully slashes the Rx latency to just 31 ms and achieves a theoretical throughput of 19.2 FPS, strictly fulfilling the constraints for real-time immersive communication.

\section{Conclusion}

We presented GS-SCNet, a 3DGS-enabled semantic multi-view coding framework for real-time immersive video communication. Leveraging cross-view parallel semantic autoencoders with disparity-aware feature interaction, GS-SCNet effectively suppresses redundancy between stereo views and supports real-time encoding and decoding. A compact Gaussian-parameter predictor then converts the decoded semantic features directly into 3D splatting attributes. By unifying the extraction and utilization of inter-view correlations, our method effectively eliminates the repeated, high-complexity geometric modeling that severely bottlenecks independent multi-view codecs and 3D reconstructors in decoupled paradigms. 

Extensive experiments demonstrate that GS-SCNet consistently surpasses both conventional and learned video codecs in compression efficiency, rendering fidelity, and perceptual quality. By shifting the optimization paradigm from Rate-Distortion to Rate-Rendering Distortion and efficiently eliminating cross-view redundancy at the semantic level, our method achieves substantially superior coding performance. Notably, cross-domain evaluations on real-world datasets reveal that our method exhibits strong generalization to unseen subjects and robustness against compression artifacts, effectively preventing the geometric misalignments and cross-view fusion artifacts that frequently degrade state-of-the-art decoupled paradigms. Furthermore, complexity analysis confirms that our unified end-to-end architecture significantly reduces computational overhead, achieving high-speed inference that satisfies the stringent latency constraints of real-time telepresence.

\bibliographystyle{IEEEtran}
\bibliography{reference}

@ARTICLE{H265,
  author={Sullivan, Gary J. and Ohm, Jens-Rainer and Han, Woo-Jin and Wiegand, Thomas},
  journal={IEEE Transactions on Circuits and Systems for Video Technology}, 
  title={Overview of the High Efficiency Video Coding {(HEVC)} Standard}, 
  year={2012},
  volume={22},
  number={12},
  pages={1649-1668},
  doi={10.1109/TCSVT.2012.2221191}}

@ARTICLE{MVHEVC,
  author={Tech, Gerhard and Chen, Ying and Müller, Karsten and Ohm, Jens-Rainer and Vetro, Anthony and Wang, Ye-Kui},
  journal={IEEE Transactions on Circuits and Systems for Video Technology}, 
  title={Overview of the Multiview and {3D} Extensions of High Efficiency Video Coding}, 
  year={2016},
  volume={26},
  number={1},
  pages={35-49},
  doi={10.1109/TCSVT.2015.2477935}}

@InProceedings{DCVCDC,
    author    = {Li, Jiahao and Li, Bin and Lu, Yan},
    title     = {Neural Video Compression With Diverse Contexts},
    booktitle = {Proceedings of the IEEE/CVF Conference on Computer Vision and Pattern Recognition (CVPR)},
    month     = {June},
    year      = {2023},
    pages     = {22616-22626}
}

@InProceedings{DVC,
author = {Lu, Guo and Ouyang, Wanli and Xu, Dong and Zhang, Xiaoyun and Cai, Chunlei and Gao, Zhiyong},
title = {{DVC}: An End-To-End Deep Video Compression Framework},
booktitle = {Proceedings of the IEEE/CVF Conference on Computer Vision and Pattern Recognition (CVPR)},
month = {June},
year = {2019}
}

@inproceedings{DCVCFM,
  title={Neural video compression with feature modulation},
  author={Li, Jiahao and Li, Bin and Lu, Yan},
  booktitle={Proceedings of the IEEE/CVF Conference on Computer Vision and Pattern Recognition},
  pages={26099--26108},
  year={2024}
}

@article{DCVCRT,
  title={Towards Practical Real-Time Neural Video Compression},
  author={Jia, Zhaoyang and Li, Bin and Li, Jiahao and Xie, Wenxuan and Qi, Linfeng and Li, Houqiang and Lu, Yan},
  journal={arXiv preprint arXiv:2502.20762},
  year={2025}
}

@article{huang2022iscom,
  title={ISCom: Interest-aware semantic communication scheme for point cloud video streaming},
  author={Huang, Yakun and Bai, Boyuan and Zhu, Yuanwei and Qiao, Xiuquan and Su, Xiang and Zhang, Ping},
  journal={arXiv preprint arXiv:2210.06808},
  year={2022}
}

@article{mildenhall2021nerf,
  title={{NeRF}: Representing scenes as neural radiance fields for view synthesis},
  author={Mildenhall, Ben and Srinivasan, Pratul P and Tancik, Matthew and Barron, Jonathan T and Ramamoorthi, Ravi and Ng, Ren},
  journal={Communications of the ACM},
  volume={65},
  number={1},
  pages={99--106},
  year={2021},
  publisher={ACM New York, NY, USA}
}

@article{kerbl20233d,
  title={{3D Gaussian Splatting} for real-time radiance field rendering.},
  author={Kerbl, Bernhard and Kopanas, Georgios and Leimk{\"u}hler, Thomas and Drettakis, George},
  journal={ACM Trans. Graph.},
  volume={42},
  number={4},
  pages={139--1},
  year={2023}
}

@inproceedings{chen2024mvsplat,
  title={{MVSplat: Efficient 3D Gaussian Splatting from sparse multi-view images}},
  author={Chen, Yuedong and Xu, Haofei and Zheng, Chuanxia and Zhuang, Bohan and Pollefeys, Marc and Geiger, Andreas and Cham, Tat-Jen and Cai, Jianfei},
  booktitle={European Conference on Computer Vision},
  pages={370--386},
  year={2024},
  organization={Springer}
}

@inproceedings{zheng2024gps,
  title={{GPS-Gaussian: Generalizable pixel-wise 3D Gaussian Splatting for real-time human novel view synthesis}},
  author={Zheng, Shunyuan and Zhou, Boyao and Shao, Ruizhi and Liu, Boning and Zhang, Shengping and Nie, Liqiang and Liu, Yebin},
  booktitle={Proceedings of the IEEE/CVF conference on computer vision and pattern recognition},
  pages={19680--19690},
  year={2024}
}

@article{tu2024tele,
  title={Tele-aloha: A low-budget and high-authenticity telepresence system using sparse {RGB} cameras},
  author={Tu, Hanzhang and Shao, Ruizhi and Dong, Xue and Zheng, Shunyuan and Zhang, Hao and Chen, Lili and Wang, Meili and Li, Wenyu and Ma, Siyan and Zhang, Shengping and others},
  journal={arXiv preprint arXiv:2405.14866},
  year={2024}
}

@incollection{starline,
  title={Project starline: A high-fidelity telepresence system},
  author={Lawrence, Jason and Overbeck, Ryan and Prives, Todd and Fortes, Tommy and Roth, Nikki and Newman, Brett},
  booktitle={ACM SIGGRAPH 2024 Emerging Technologies},
  pages={1--2},
  year={2024}
}

@inproceedings{chen2022lsvc,
  title={{LSVC}: A learning-based stereo video compression framework},
  author={Chen, Zhenghao and Lu, Guo and Hu, Zhihao and Liu, Shan and Jiang, Wei and Xu, Dong},
  booktitle={Proceedings of the IEEE/CVF Conference on Computer Vision and Pattern Recognition},
  pages={6073--6082},
  year={2022}
}

@inproceedings{charatan2024pixelsplat,
  title={{pixelSplat}: {3D Gaussian Splats} from image pairs for scalable generalizable 3{D} reconstruction},
  author={Charatan, David and Li, Sizhe Lester and Tagliasacchi, Andrea and Sitzmann, Vincent},
  booktitle={Proceedings of the IEEE/CVF conference on computer vision and pattern recognition},
  pages={19457--19467},
  year={2024}
}

@article{balle2016end,
  title={End-to-end optimized image compression},
  author={Ball{\'e}, Johannes and Laparra, Valero and Simoncelli, Eero P},
  journal={arXiv preprint arXiv:1611.01704},
  year={2016}
}

@article{balle2018variational,
  title={Variational image compression with a scale hyperprior},
  author={Ball{\'e}, Johannes and Minnen, David and Singh, Saurabh and Hwang, Sung Jin and Johnston, Nick},
  journal={arXiv preprint arXiv:1802.01436},
  year={2018}
}

@article{minnen2018joint,
  title={Joint autoregressive and hierarchical priors for learned image compression},
  author={Minnen, David and Ball{\'e}, Johannes and Toderici, George D},
  journal={Advances in neural information processing systems},
  volume={31},
  year={2018}
}

@inproceedings{hu2020coarse,
  title={Coarse-to-fine hyper-prior modeling for learned image compression},
  author={Hu, Yueyu and Yang, Wenhan and Liu, Jiaying},
  booktitle={Proceedings of the AAAI Conference on Artificial Intelligence},
  volume={34},
  number={07},
  pages={11013--11020},
  year={2020}
}

@inproceedings{he2022elic,
  title={{ELIC}: Efficient learned image compression with unevenly grouped space-channel contextual adaptive coding},
  author={He, Dailan and Yang, Ziming and Peng, Weikun and Ma, Rui and Qin, Hongwei and Wang, Yan},
  booktitle={Proceedings of the IEEE/CVF Conference on Computer Vision and Pattern Recognition},
  pages={5718--5727},
  year={2022}
}

@article{ascenso2023jpeg,
  title={The {JPEG AI} standard: Providing efficient human and machine visual data consumption},
  author={Ascenso, Jo{\~a}o and Alshina, Elena and Ebrahimi, Touradj},
  journal={Ieee Multimedia},
  volume={30},
  number={1},
  pages={100--111},
  year={2023},
  publisher={IEEE}
}

@inproceedings{rippel2021elf,
  title={{ELF-VC}: Efficient learned flexible-rate video coding},
  author={Rippel, Oren and Anderson, Alexander G and Tatwawadi, Kedar and Nair, Sanjay and Lytle, Craig and Bourdev, Lubomir},
  booktitle={Proceedings of the IEEE/CVF International Conference on Computer Vision},
  pages={14479--14488},
  year={2021}
}

@inproceedings{pourreza2023boosting,
  title={Boosting neural video codecs by exploiting hierarchical redundancy},
  author={Pourreza, Reza and Le, Hoang and Said, Amir and Sautiere, Guillaume and Wiggers, Auke},
  booktitle={Proceedings of the IEEE/CVF Winter Conference on Applications of Computer Vision},
  pages={5355--5364},
  year={2023}
}

@inproceedings{hu2021fvc,
  title={{FVC}: A new framework towards deep video compression in feature space},
  author={Hu, Zhihao and Lu, Guo and Xu, Dong},
  booktitle={Proceedings of the IEEE/CVF conference on computer vision and pattern recognition},
  pages={1502--1511},
  year={2021}
}

@article{li2021deep,
  title={Deep contextual video compression},
  author={Li, Jiahao and Li, Bin and Lu, Yan},
  journal={Advances in Neural Information Processing Systems},
  volume={34},
  pages={18114--18125},
  year={2021}
}

@article{sheng2022temporal,
  title={Temporal context mining for learned video compression},
  author={Sheng, Xihua and Li, Jiahao and Li, Bin and Li, Li and Liu, Dong and Lu, Yan},
  journal={IEEE Transactions on Multimedia},
  volume={25},
  pages={7311--7322},
  year={2022},
  publisher={IEEE}
}

@inproceedings{li2022hybrid,
  title={Hybrid spatial-temporal entropy modelling for neural video compression},
  author={Li, Jiahao and Li, Bin and Lu, Yan},
  booktitle={Proceedings of the 30th ACM International Conference on Multimedia},
  pages={1503--1511},
  year={2022}
}

@inproceedings{tian2023towards,
  title={Towards real-time neural video codec for cross-platform application using calibration information},
  author={Tian, Kuan and Guan, Yonghang and Xiang, Jinxi and Zhang, Jun and Han, Xiao and Yang, Wei},
  booktitle={Proceedings of the 31st ACM International Conference on Multimedia},
  pages={7961--7970},
  year={2023}
}

@article{perkins2002data,
  title={Data compression of stereopairs},
  author={Perkins, Michael G},
  journal={IEEE Transactions on communications},
  volume={40},
  number={4},
  pages={684--696},
  year={2002},
  publisher={IEEE}
}

@ARTICLE{MVC,
  author={Vetro, Anthony and Wiegand, Thomas and Sullivan, Gary J.},
  journal={Proceedings of the IEEE}, 
  title={Overview of the Stereo and Multiview Video Coding Extensions of the {H.264/MPEG-4 AVC} Standard}, 
  year={2011},
  volume={99},
  number={4},
  pages={626-642},
  doi={10.1109/JPROC.2010.2098830}}

@inproceedings{hou2024llss,
  title={Low-latency neural stereo streaming},
  author={Hou, Qiqi and Farhadzadeh, Farzad and Said, Amir and Sautiere, Guillaume and Le, Hoang},
  booktitle={Proceedings of the IEEE/CVF Conference on Computer Vision and Pattern Recognition},
  pages={7974--7984},
  year={2024}
}

@article{zhou2025gps,
  title={{GPS-Gaussian+: Generalizable pixel-wise 3D Gaussian Splatting for real-time human-scene rendering from sparse views}},
  author={Zhou, Boyao and Zheng, Shunyuan and Tu, Hanzhang and Shao, Ruizhi and Liu, Boning and Zhang, Shengping and Nie, Liqiang and Liu, Yebin},
  journal={IEEE Transactions on Pattern Analysis and Machine Intelligence},
  year={2025},
  publisher={IEEE}
}

@inproceedings{szymanowicz2024splatter,
  title={{Splatter image: Ultra-fast single-view 3D reconstruction}},
  author={Szymanowicz, Stanislaw and Rupprecht, Chrisitian and Vedaldi, Andrea},
  booktitle={Proceedings of the IEEE/CVF conference on computer vision and pattern recognition},
  pages={10208--10217},
  year={2024}
}

@inproceedings{lipson2021raft,
  title={Raft-stereo: Multilevel recurrent field transforms for stereo matching},
  author={Lipson, Lahav and Teed, Zachary and Deng, Jia},
  booktitle={2021 International Conference on 3D Vision (3DV)},
  pages={218--227},
  year={2021},
  organization={IEEE}
}

@inproceedings{wang20244d,
  title={{4D-DRESS: A 4D dataset of real-world human clothing with semantic annotations}},
  author={Wang, Wenbo and Ho, Hsuan-I and Guo, Chen and Rong, Boxiang and Grigorev, Artur and Song, Jie and Zarate, Juan Jose and Hilliges, Otmar},
  booktitle={Proceedings of the IEEE/CVF Conference on Computer Vision and Pattern Recognition},
  pages={550--560},
  year={2024}
}

@inproceedings{shen2023x,
  title={{X-Avatar: Expressive human avatars}},
  author={Shen, Kaiyue and Guo, Chen and Kaufmann, Manuel and Zarate, Juan Jose and Valentin, Julien and Song, Jie and Hilliges, Otmar},
  booktitle={Proceedings of the IEEE/CVF Conference on Computer Vision and Pattern Recognition},
  pages={16911--16921},
  year={2023}
}

@article{bjontegaard2001calculation,
  title={{Calculation of average PSNR differences between RD-curves}},
  author={Bjontegaard, Gisle},
  journal={ITU SG16 Doc. VCEG-M33},
  year={2001}
}

@inproceedings{chollet2017xception,
  title={{Xception: Deep learning with depthwise separable convolutions}},
  author={Chollet, Fran{\c{c}}ois},
  booktitle={Proceedings of the IEEE conference on computer vision and pattern recognition},
  pages={1251--1258},
  year={2017}
}

@inproceedings{DX,
author = {Yang, Dingxi and Qin, Zhijin},
title = {Generalizable 3D Gaussian Splatting Enabled Immersive Video Communications with Semantic Coding},
year = {2025},
isbn = {9798400719974},
publisher = {Association for Computing Machinery},
address = {New York, NY, USA},
url = {https://doi.org/10.1145/3742889.3768350},
doi = {10.1145/3742889.3768350},
booktitle = {Proceedings of the 3rd ACM Workshop on Mobile Immersive Computing, Networking, and Systems},
pages = {36–42},
numpages = {7},
location = {Hong Kong, China},
series = {ImmerCom '25}
}

@inproceedings{tang2025hgc,
  title={HGC-Avatar: Hierarchical Gaussian Compression for Streamable Dynamic 3D Avatars},
  author={Tang, Haocheng and Yan, Ruoke and Yin, Xinhui and Zhang, Qi and Zhang, Xinfeng and Ma, Siwei and Gao, Wen and Jia, Chuanmin},
  booktitle={Proceedings of the 33rd ACM International Conference on Multimedia},
  pages={8125--8134},
  year={2025}
}

@article{hu2024eva,
  title={EVA-Gaussian: 3D Gaussian-based Real-time Human Novel View Synthesis under Diverse Multi-view Camera Settings},
  author={Hu, Yingdong and Liu, Zhening and Shao, Jiawei and Lin, Zehong and Zhang, Jun},
  journal={arXiv preprint arXiv:2410.01425},
  year={2024}
}

@inproceedings{cheng2023dna,
  title={Dna-rendering: A diverse neural actor repository for high-fidelity human-centric rendering},
  author={Cheng, Wei and Chen, Ruixiang and Fan, Siming and Yin, Wanqi and Chen, Keyu and Cai, Zhongang and Wang, Jingbo and Gao, Yang and Yu, Zhengming and Lin, Zhengyu and others},
  booktitle={Proceedings of the IEEE/CVF International Conference on Computer Vision},
  pages={19982--19993},
  year={2023}
}

@inproceedings{kwon2024generalizable,
  title={Generalizable human gaussians for sparse view synthesis},
  author={Kwon, Youngjoong and Fang, Baole and Lu, Yixing and Dong, Haoye and Zhang, Cheng and Carrasco, Francisco Vicente and Mosella-Montoro, Albert and Xu, Jianjin and Takagi, Shingo and Kim, Daeil and others},
  booktitle={European Conference on Computer Vision},
  pages={451--468},
  year={2024},
  organization={Springer}
}

@inproceedings{chen2024hac,
  title={Hac: Hash-grid assisted context for 3d gaussian splatting compression},
  author={Chen, Yihang and Wu, Qianyi and Lin, Weiyao and Harandi, Mehrtash and Cai, Jianfei},
  booktitle={European Conference on Computer Vision},
  pages={422--438},
  year={2024},
  organization={Springer}
}

@article{lu2024multi,
  title={Multi-view image-based 3d reconstruction in indoor scenes: a survey},
  author={LU, Ping and SHI, Wenzhe and QIAO, Xiuquan},
  journal={ZTE Communications},
  volume={22},
  number={3},
  pages={91},
  year={2024}
}

\begin{IEEEbiography}
[{\includegraphics[width=1in,height=1.25in,clip,keepaspectratio]{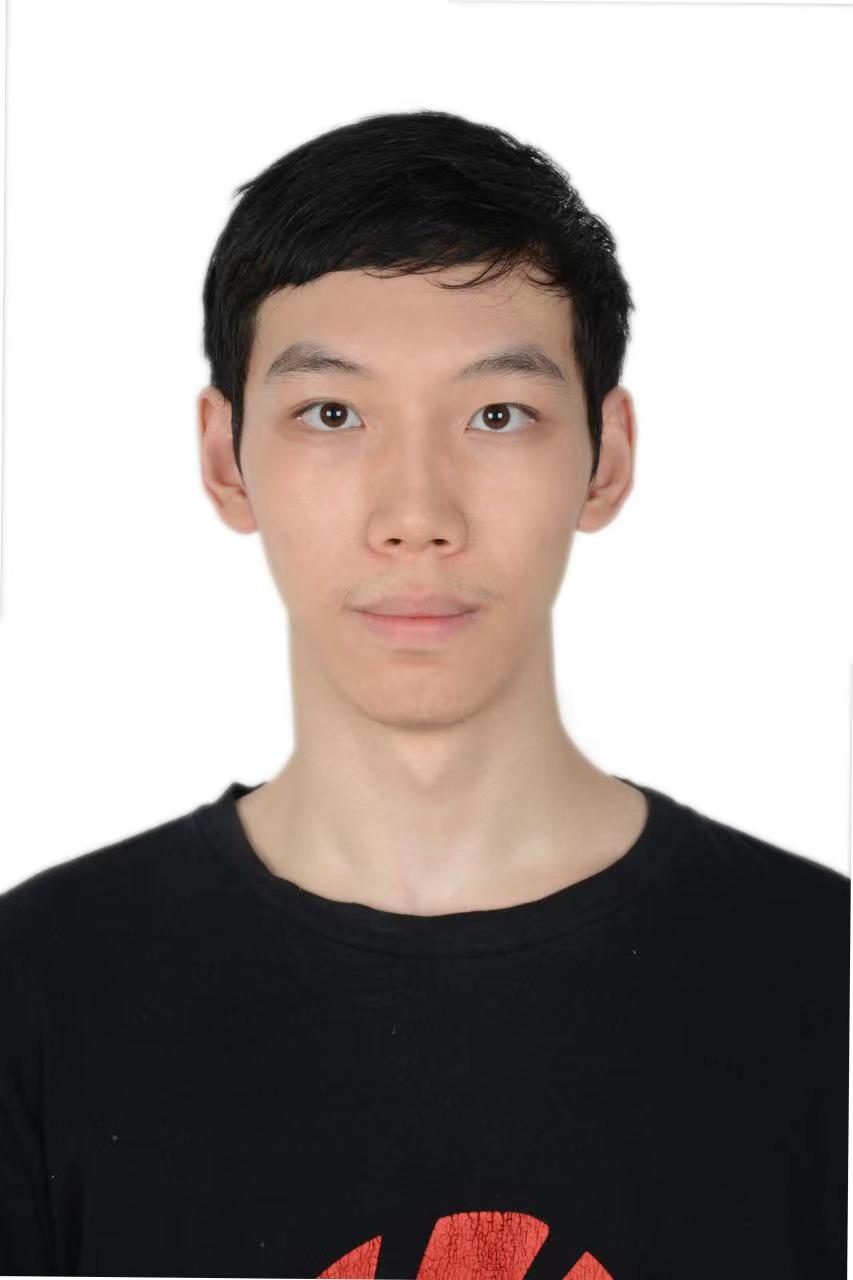}}]{Dingxi Yang} (Student Member, IEEE) received the bachelor’s degree in electronic engineering from Tsinghua University, Beijing, China, in 2023, where he is currently pursuing the Ph.D. degree with the Department of Electronic Engineering. His research interests include immersive communication and semantic coding.
\end{IEEEbiography}

\begin{IEEEbiography}
[{\includegraphics[width=1in,height=1.25in,clip,keepaspectratio]{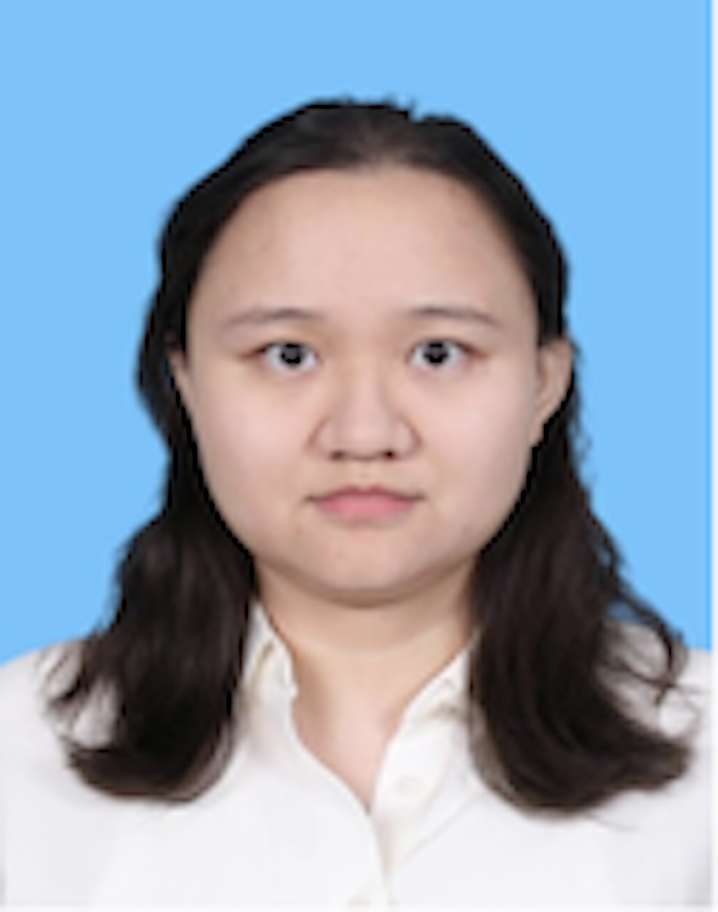}}]{Wenqi Guo}  received the B.S. degree from the Department of Software Engineering, University of Electronic Science and Technology of China (UESTC), Chengdu, China, in 2020, and the Ph.D. degree from the School of Computer Science, Shanghai Jiao Tong University (SJTU), Shanghai, China, in 2026. She is currently a Post-Doctoral Fellow with the Department of Automation, Tsinghua University, Beijing, China. Her current research interests include graph neural networks, graph matching, and self-supervised learning.
\end{IEEEbiography}

\begin{IEEEbiography}
[{\includegraphics[width=1in,height=1.25in,clip,keepaspectratio]{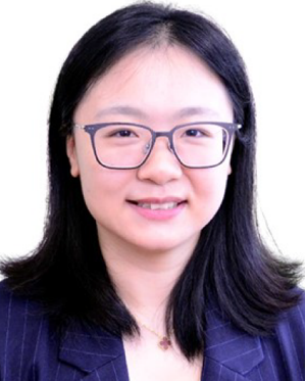}}]{Yue Liu} 
(Member, IEEE) received the B.Sc. degree from Beijing University of Posts and Telecommunications, Beijing, China, in 2010, and the Ph.D. degree in electronic engineering from the Queen Mary University of London, London, U.K., in 2014. She joined the MPI-QMUL Information Systems Research Centre, Macau, China, in 2014, as a Researcher and has been a Lecturer with the Faculty of Applied Sciences, Macao Polytechnic University, Macau, since 2015. Her research interests include radio resource management, wireless human sensing, deep reinforcement learning, and AI-driven wireless networks. She is currently serves as a reviewer for IEEE COMMUNICATIONS LETTERS, IEEE TRANSACTIONS ON LEARNING TECHNOLOGIES, and IEEE VTC, and the Symposium Chair for IEEE ICSPS.
\end{IEEEbiography}

\begin{IEEEbiography}
[{\includegraphics[width=1in,height=1.25in,clip,keepaspectratio]{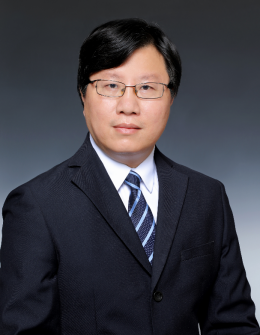}}]{Jungong Han} (Senior Member, IEEE) is currently a chair professor at the Department of Computer Science, Aberystwyth University, U.K. He also holds an honorary professorship with the University of Warwick, U.K. He served as a faculty member (2017-2019) at Lancaster University, U.K. His research interests include computer vision, artificial intelligence and machine learning. He has published more than 60 IEEE/ACM Transactions papers and 50 A* conference papers.
\end{IEEEbiography}

\begin{IEEEbiography}
[{\includegraphics[width=1in,height=1.25in,clip,keepaspectratio]{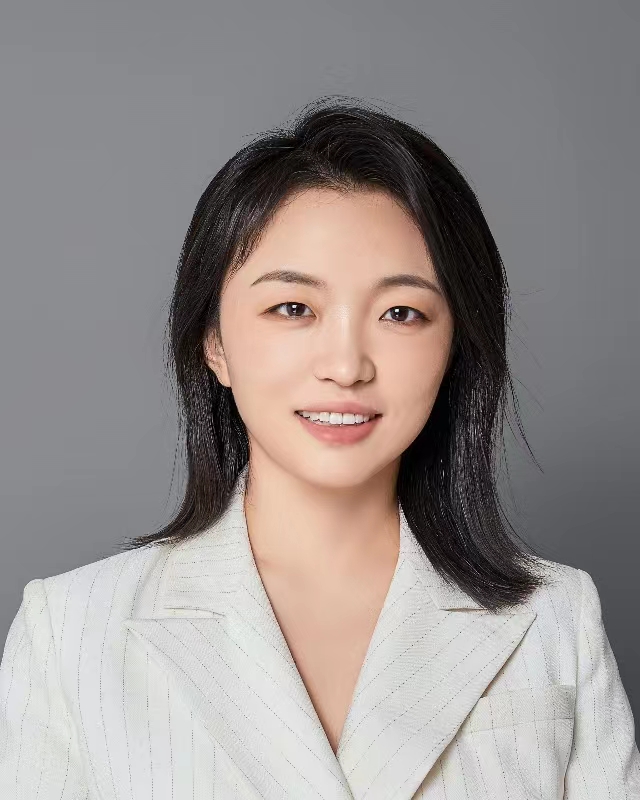}}]{Zhijin Qin} (Senior Member, IEEE) is currently an Associate Professor at the Department of Electronic Engineering, Tsinghua University, Beijing, China. Her research interests include semantic communications and sparse signal processing. She has received several awards, including the 2017 IEEE GLOBECOM Best Paper Award, the 2018 IEEE Signal Processing Society Young Author Best Paper Award, the 2021 IEEE Communications Society SPCC Early Achievement Award, and the 2022 IEEE Communications Society Fred W. Ellersick Prize. She served as the Guest Editor for the Special Issue on Semantic Communications of IEEE Journal on Selected Areas in Communications and as an Associate Editor for IEEE Transactions on Communications. 
\end{IEEEbiography}

\end{document}